\documentclass[conference,a4paper,10pt]{IEEEtran}
 \usepackage{mathtools}
 \usepackage{amsfonts}
 \usepackage{amsmath}
 \usepackage{bbm}
 \usepackage{amsthm}
 \usepackage{amssymb}

\usepackage{tablefootnote}
\usepackage{url}

\usepackage{algorithm}
\usepackage{algorithmicx}
\usepackage{algpseudocode}
\usepackage{booktabs}
\usepackage{textcomp}
\usepackage{enumitem}
\usepackage{subfigure}
\usepackage{graphicx,epstopdf}
\usepackage{xcolor}
\usepackage{url}
\usepackage{cite}
\usepackage{soul}
\usepackage{balance}
\usepackage{multirow}

\definecolor{hgreen}{rgb}{0, 0.5, 0}

\ifCLASSINFOpdf
\else
\fi
\begin{document}

%
\title{Site-Specific and Frequency-Dependent Channel Characterization and MIMO Performance in FR3}
\author{\IEEEauthorblockN{Zhuangzhuang~Cui\IEEEauthorrefmark{1}, Rudranil Chattopadhyay\IEEEauthorrefmark{1}, Emiel Vanspranghels\IEEEauthorrefmark{1}, and Sofie Pollin\IEEEauthorrefmark{1}\IEEEauthorrefmark{2}}
\IEEEauthorblockA{
\IEEEauthorrefmark{1}WaveCoRE, Department of Electrical Engineering (ESAT), KU Leuven, Belgium\\
\IEEEauthorrefmark{2}Interuniversity Microelectronics Centre (imec), Leuven, Belgium\\}
Email: \texttt{\{firstname.lastname\}@kuleuven.be}}

%



\maketitle

\begin{abstract}
Next-generation wireless systems aim to enable on-demand connectivity through dynamic spectrum utilization. Motivated by this vision, this paper investigates the propagation characteristics and MIMO performance of the upper mid-band, spanning approximately 7–24 GHz and unofficially referred to as FR3. Using site-specific ray-tracing (RT) simulations based on the Sionna framework, we analyze indoor and outdoor environments at representative frequencies across FR1, FR3, and FR2, including 3.5, 7, 10, 14, 20, 24, and 28~GHz, under both single-antenna and multi-antenna configurations. The results show that FR3 exhibits intermediate propagation behavior between sub-6~GHz and millimeter-wave bands while sustaining effective spatial multiplexing and favorable spectral efficiency. Furthermore, large-array analysis indicates that performance gains in FR3 are closely tied to antenna scaling, highlighting the importance of large-size or large-aperture MIMO architectures for practical deployments.
\end{abstract}


\section{Introduction}

Spectrum availability and propagation characteristics play a central role in determining the performance of wireless communication systems. Looking ahead to sixth-generation (6G) networks, a key objective is to enable on-demand connectivity through flexible and dynamic spectrum utilization. Achieving this goal requires a comprehensive understanding of radio propagation and system performance across a wide range of frequency bands.

Current fifth-generation (5G) networks support multi-band operation by combining sub-6~GHz and millimeter-wave (mmWave) bands, denoted as frequency ranges (FR)~1 and~2 in the 3GPP framework. While FR1 provides reliable coverage, it is constrained by limited available bandwidth. In contrast, FR2 offers large contiguous spectrum resources but suffers from severe path loss, increased channel sparsity, and high sensitivity to blockage. These limitations motivate the exploration of intermediate frequency bands that can balance bandwidth availability and propagation robustness.

Recently, increasing attention has been directed toward the upper mid-band, unofficially referred to as FR3, spanning approximately 7--24~GHz \cite{cui25dyspan}. This frequency range is expected to offer wider bandwidth than FR1 while avoiding the extreme propagation impairments typical of FR2, making it a promising candidate for future 6G deployments. Consequently, a growing body of work has investigated FR3 channel characteristics through measurement campaigns and statistical modeling.

Early studies reported channel measurements in both outdoor and indoor environments at selected FR3 frequencies, such as 6.75~GHz and 16.95~GHz \cite{fr3_mea_outdoor,fr3_mea_indoor}, and proposed path-loss models based on the close-in (CI) formulation. These works observed decreasing trends in delay and angular spreads with increasing frequency within the considered environments. While these results provide valuable insights into FR3 propagation behavior, the limited number of frequency points restricts the generality of the observed trends. Subsequent measurements at 10~GHz \cite{guo2025directional} demonstrated that angular spread strongly depends on the employed antenna array, highlighting the interaction between array geometry and spatial channel characteristics. Additional measurement campaigns have been conducted across various FR3 frequencies and environments, including suburban campuses \cite{fr3_dum}, indoor spaces such as open offices and corridors \cite{6gic_1218}, lecture halls with MIMO setups \cite{vienna_fr3}, and industrial scenarios \cite{oulu-fr3}. A comprehensive overview of FR3 channel measurements and characterization can be found in \cite{zhang_bupt}. However, most existing studies focus primarily on channel statistics and lack systematic comparisons with commonly deployed FR1 and FR2 bands. Moreover, channel characterization is often decoupled from system-level performance evaluation, limiting insights into the practical implications for MIMO deployments.

Site-specific ray-tracing techniques provide a powerful and complementary approach for studying radio propagation across multiple frequency bands and diverse environments. By incorporating detailed three-dimensional (3D) geometry and material properties derived from open-source data, such as OpenStreetMap, ray tracing enables consistent and controlled evaluation of frequency-dependent channel behavior. From a system-design perspective, assessing the achievable MIMO performance based on site-specific channels is a critical step toward practical deployment of FR3 systems. In this context, it is essential to consider both indoor environments, characterized by confined spaces and rich scattering, and outdoor urban environments, where building geometry and street layout strongly influence propagation mechanisms.

In this paper, we address two key aspects of FR3 propagation and performance: \emph{site specificity} and \emph{frequency dependency}. Using the Sionna ray-tracing framework, we generate extensive channel data in both the time and frequency domains, including channel impulse responses (CIRs) and channel frequency responses (CFRs). Based on this dataset, we analyze both propagation characteristics and MIMO performance across FR1, FR3, and FR2 bands, and provide insights into realistic FR3 deployment strategies. The main contributions of this work are summarized as follows:
\begin{itemize}
    \item We construct two representative site-specific scenarios, including an indoor laboratory and an outdoor dense urban environment extracted from Paris, and perform ray-tracing simulations across carrier frequencies ranging from 3.5~GHz to 28~GHz. The resulting channel characteristics, including Rician K-factor, delay spread, and angular spread, are shown both frequency- and site-dependency.
    \item We evaluate key MIMO performance metrics, including channel rank, condition number, and spectral efficiency (SE), and reveal distinct performance trends across frequencies and environments. These trends are explained by the interplay between multipath propagation and antenna-array responses.
    \item We investigate extremely large MIMO (XL-MIMO) regimes in FR3 and demonstrate that channel hardening is primarily governed by the antenna number, and varies across carrier frequencies. Under fair comparisons in terms of antenna count and physical array aperture, FR3 is shown to offer favorable performance relative to FR1 and FR2.
\end{itemize}

The remainder of this paper is organized as follows. Section~II introduces the simulation scenarios, system configuration, and data collection methodology. Section~III analyzes site-specific propagation characteristics across frequency bands. Section~IV evaluates MIMO performance based on frequency-domain channel responses. Section~V discusses deployment insights for FR3 with a focus on antenna scaling in MIMO systems. Finally, Section~VI concludes the paper.

\section{Site-Specific Channel Simulations}

To investigate the propagation characteristics and MIMO performance across different frequency bands, we consider two representative scenarios, namely an indoor laboratory environment and an outdoor urban environment, as shown in Fig.~\ref{env_sites}. Ray-tracing simulations are conducted using the open-source Sionna RT framework, as described in \cite{aoudia2025sionnart}, which enables site-specific channel modeling with physics-based electromagnetic propagation mechanisms. In this section, we introduce the simulated scenarios, the frequency and antenna configurations, and the channel data collection procedure.

\subsection{Simulated Scenarios}
\subsubsection{Indoor Scenario}
The indoor scenario shown in Fig.~\ref{indoor} represents a laboratory-sized room equipped with typical furniture and electronic equipment, forming a rich-scattering environment commonly encountered in indoor wireless deployments. The room dimensions are $7.6 \times 10.5 \times 3.1$~m$^3$, and the environment is modeled as an enclosed space comprising walls, floor, and ceiling ($x$ and $y$ coordinates in [0, 7.6]~m and [0, 10.5]~m). The electromagnetic material properties are adopted from \cite{aoudia2025sionnart} and include glass, plasterboard, wood, metal, plywood, ceiling board, and concrete. Most of these materials exhibit weak frequency dependence over the 1--40~GHz range. The transmitter (Tx) is fixed, and the receiver (Rx) locations are sampled 100 times during simulation, allowing for the presence of both line-of-sight (LoS) and non-LoS (NLoS) components. 

\subsubsection{Outdoor Scenario}
\begin{figure}[!t]
  \centering
   \subfigure[Indoor (lab)]{\includegraphics[width=3in]{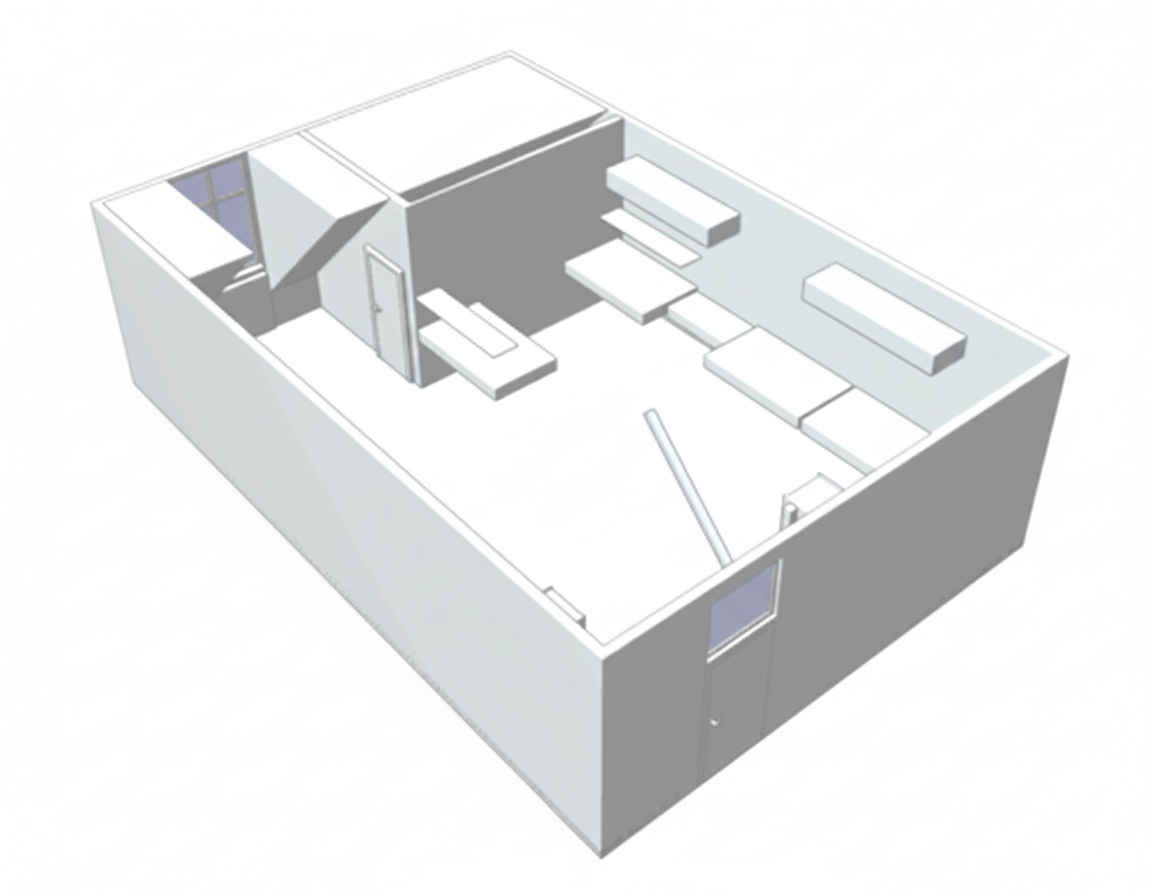}\label{indoor}}
   \subfigure[Outdoor (urban city)]{\includegraphics[width=3in]{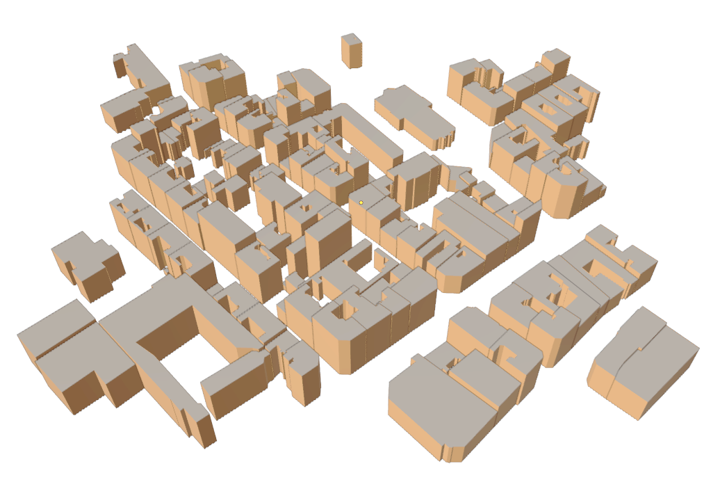}\label{outdoor}}
  \caption{Two representative scenarios considered in the ray-tracing simulations.} 
  \label{env_sites}
 \end{figure}

The outdoor scenario shown in Fig.~\ref{outdoor} is extracted from a dense urban area in Paris using OpenStreetMap data. The simulated region spans longitude and latitude ranges of [2.2962$^\circ$,~2.2995$^\circ$] and [48.8380$^\circ$,~48.8399$^\circ$], respectively, corresponding to an area of approximately $300 \times 300$~m$^2$ with buildings of varying heights ($xy$ coordinate in [-150, 150]~m). The dominant scatterers in this environment are buildings and the ground surface, forming a typical dense urban setting characterized by street canyons and complex propagation mechanisms. These include building-induced reflections, diffractions, and shadowing effects. The building geometries are imported into the Sionna RT framework and assigned realistic electromagnetic material properties, such as brick, concrete, and marble. The Tx is placed at a height representative of a base-station deployment (e.g., 30~m), while the Rx is positioned at pedestrian height (1.5~m), reflecting a typical outdoor user scenario.

\subsection{Frequency Bands and Antenna Arrays}

We consider seven representative carrier frequencies: 3.5~GHz and 28~GHz, corresponding to deployed 5G FR1 and FR2 bands, respectively, as well as five candidate FR3 frequencies at 7~GHz, 10~GHz, 14~GHz, 20~GHz, and 24~GHz. These frequencies are selected to span FR3 while enabling direct comparison with existing deployments.
For each carrier frequency, the system bandwidth is chosen based on practical considerations and spectrum availability, and is consistent with typical deployments in the corresponding frequency bands.

Both single-antenna and multi-antenna configurations are considered. For MIMO evaluations, uniform planar arrays (UPAs) with half-wavelength inter-element spacing are employed. Unless otherwise specified, a $2 \times 2$ UPA is used at both the Tx and Rx as a baseline configuration. To investigate the transition from conventional MIMO to XL-MIMO, a range of array sizes is examined. All antenna elements follow the specifications defined in 3GPP T.R.~38.901~\cite{3gpp38901}, and vertical polarization is assumed at both the Tx and Rx.

\begin{figure}[!t]
\centering
\subfigure[][Indoor]{
   \includegraphics[width=1.6in]{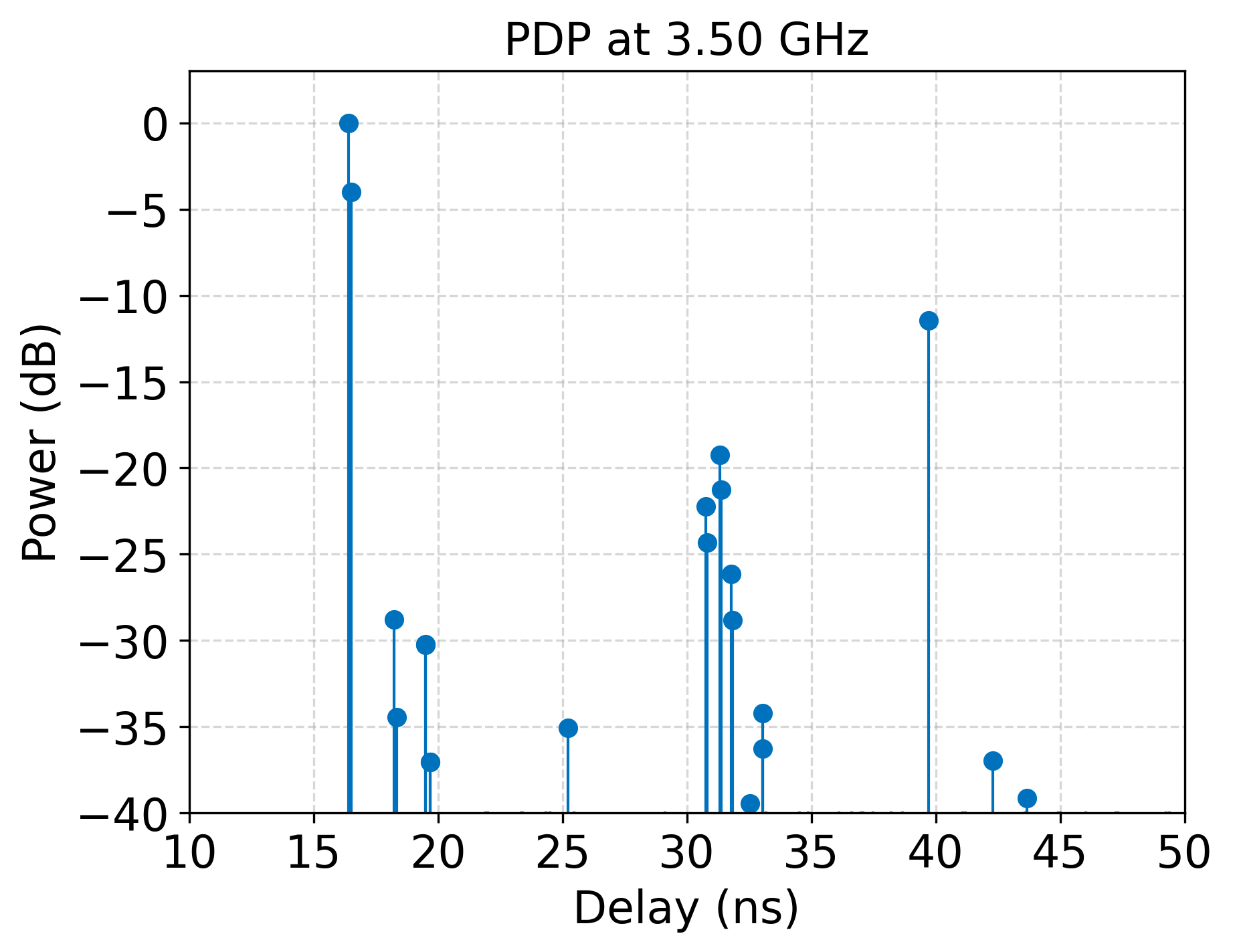}}
 \subfigure[][Indoor]{
    \includegraphics[width=1.6in]{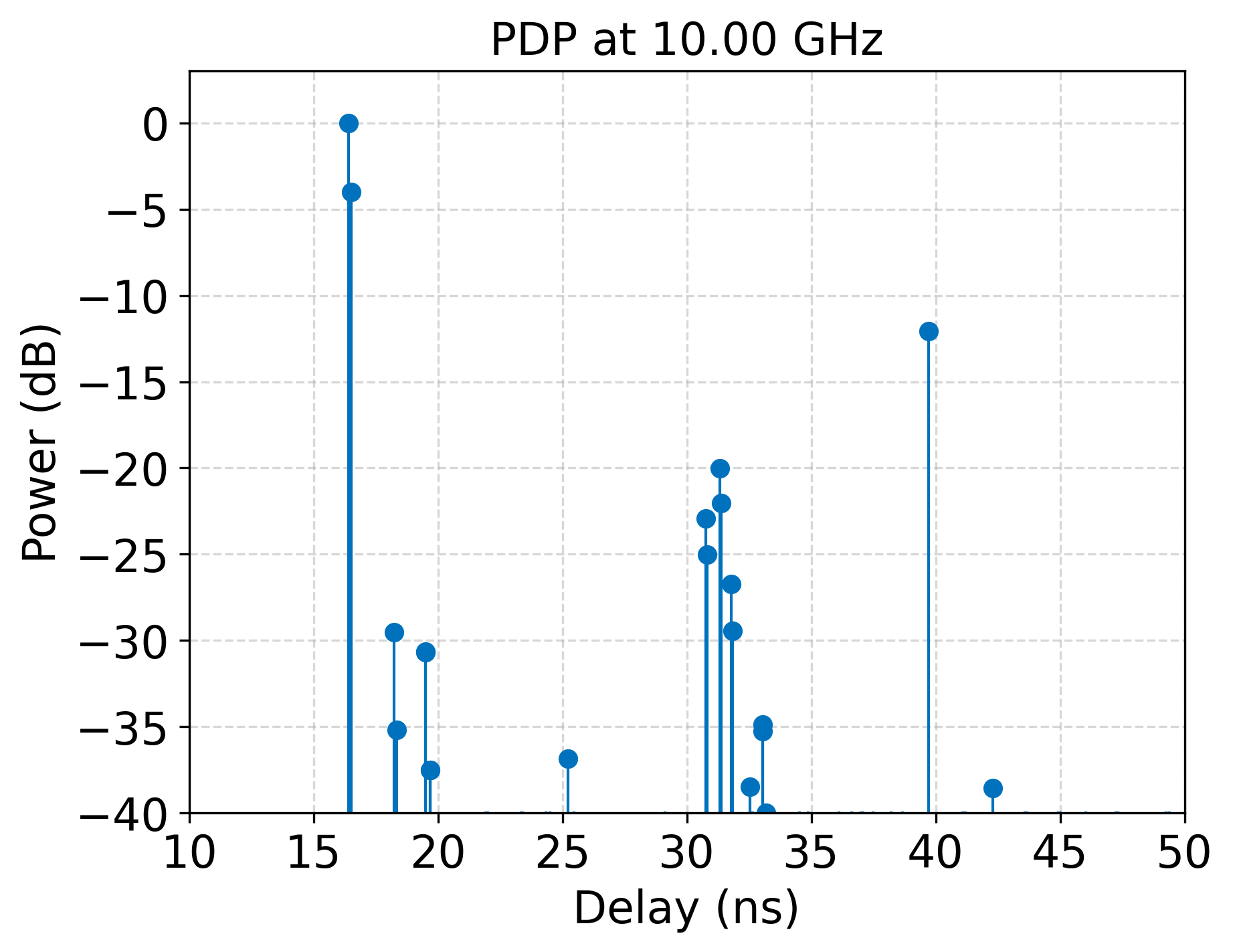}}
    \subfigure[][Outdoor]{
    \includegraphics[width=1.6in]{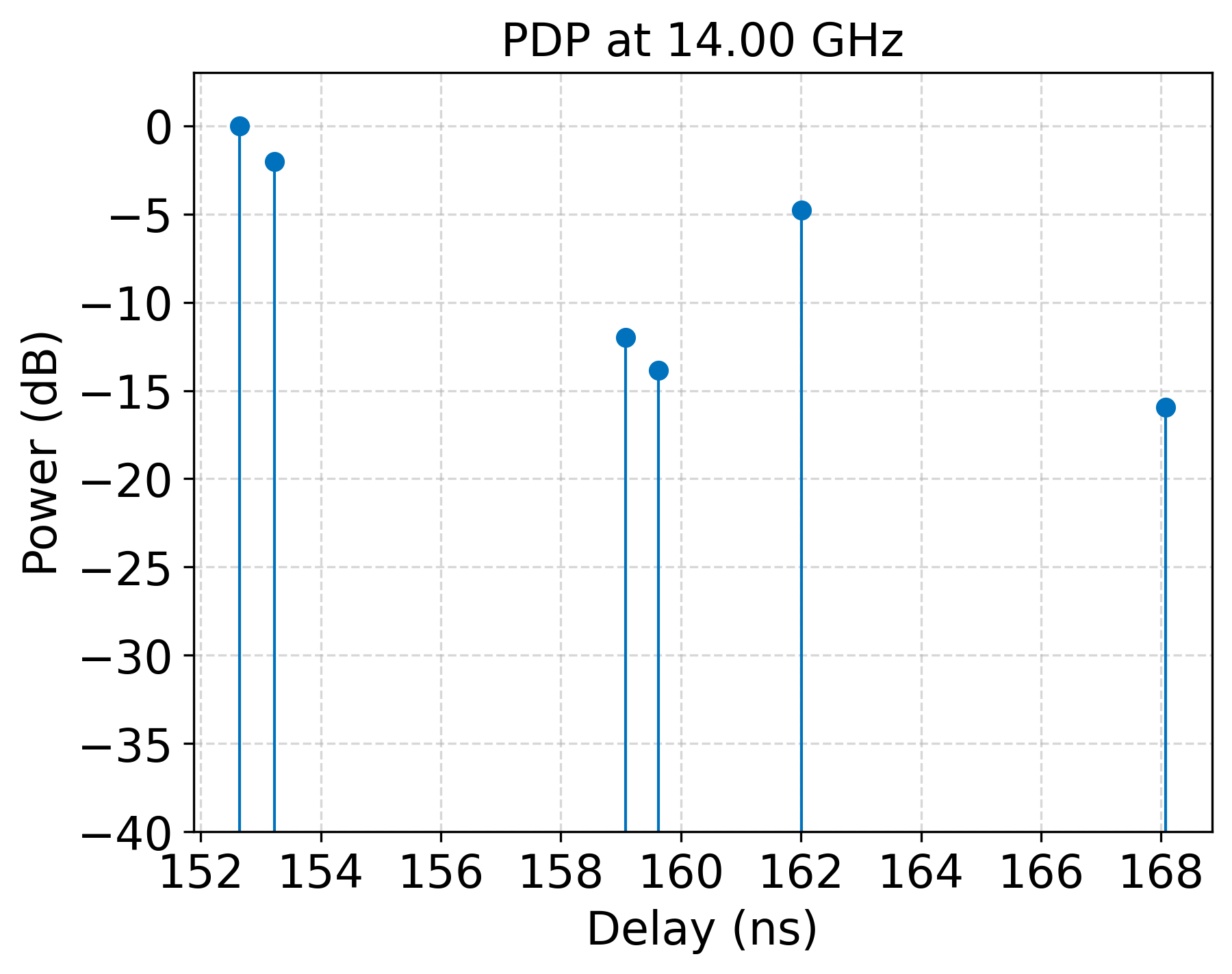}}
    \subfigure[][Outdoor]{
    \includegraphics[width=1.6in]{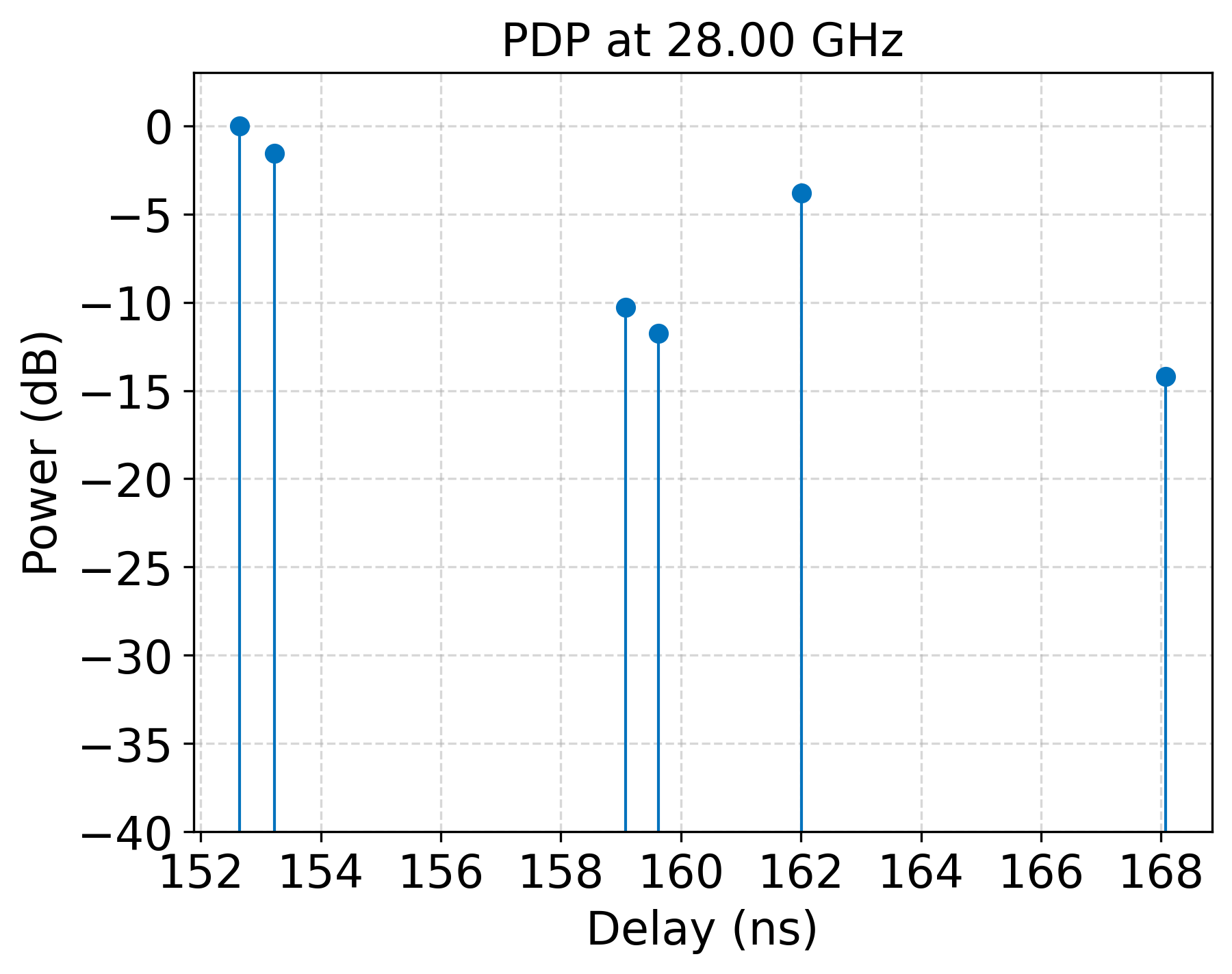}}
\caption{PDP examples for indoor and outdoor at different frequencies.}
\label{pdp}
\end{figure}

\subsection{Ray-Tracing and Channel Data Collection}
Ray-tracing simulations are conducted using Sionna RT, accounting for LoS, specular reflection, diffraction, and scattering mechanisms. For each Tx-Rx configuration and carrier frequency, the CIR is generated by aggregating all valid propagation paths within a predefined dynamic range of power (40 dB by default) and reflection order. Specifically, up to fifth-order and third-order reflections are considered for the indoor and outdoor environments, respectively, given that $> 3$rd-order reflections outdoors are generally out of the dynamic range. Based on the obtained CIRs, key propagation metrics, including the Rician K-factor, root-mean-square delay spread (RMS-DS), and angular spreads, are subsequently extracted.
The CFRs over a set of subcarriers determined by the target system bandwidth and subcarrier spacing are also collected. The resulting CFRs are used to evaluate MIMO performance metrics, including channel rank, condition number, and SE.

To ensure statistical reliability while preserving the site-specific characteristics inherent to ray-tracing-based modeling, all results are averaged over multiple channel realizations. For each site, the CIR and CFR are location-specific and correspond to individual Tx-Rx pairs. To capture site-specific channel behavior, 100 Rx locations are randomly generated within the sites. This data collection process is repeated for every carrier frequency and simulation setup, including different antenna array configurations and bandwidths, resulting in a site-specific and frequency-dependent MIMO dataset with dimensions: ${\rm Sites} \times {\rm Locations} \times f_c \times {\rm BW} \times {\rm MIMO}$.

\begin{figure}[!t]
\centering
\subfigure[][Indoor]{
   \includegraphics[width=3in]{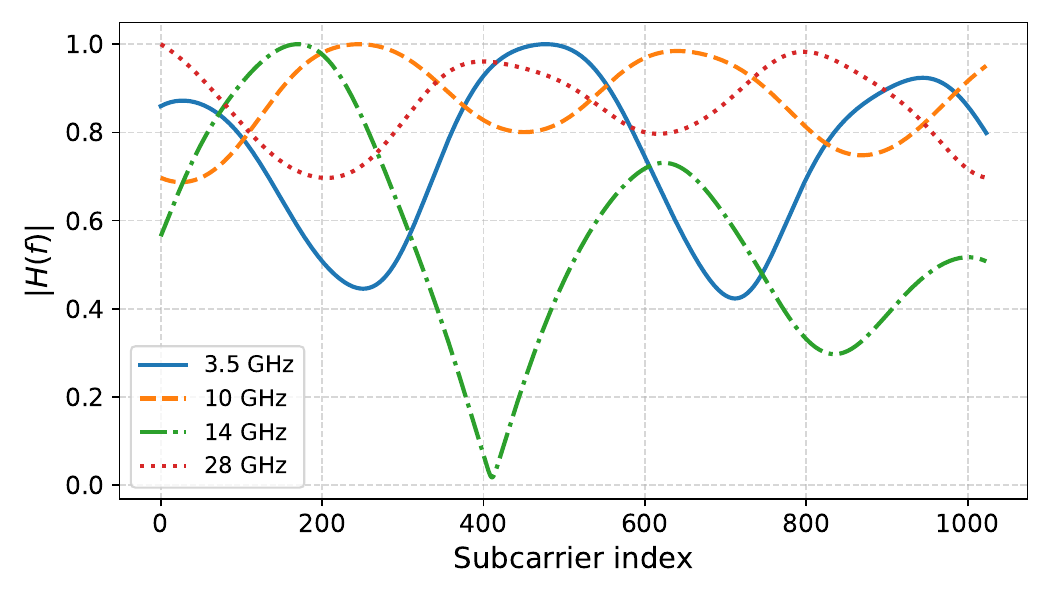}}
 \subfigure[][Outdoor]{
    \includegraphics[width=3in]{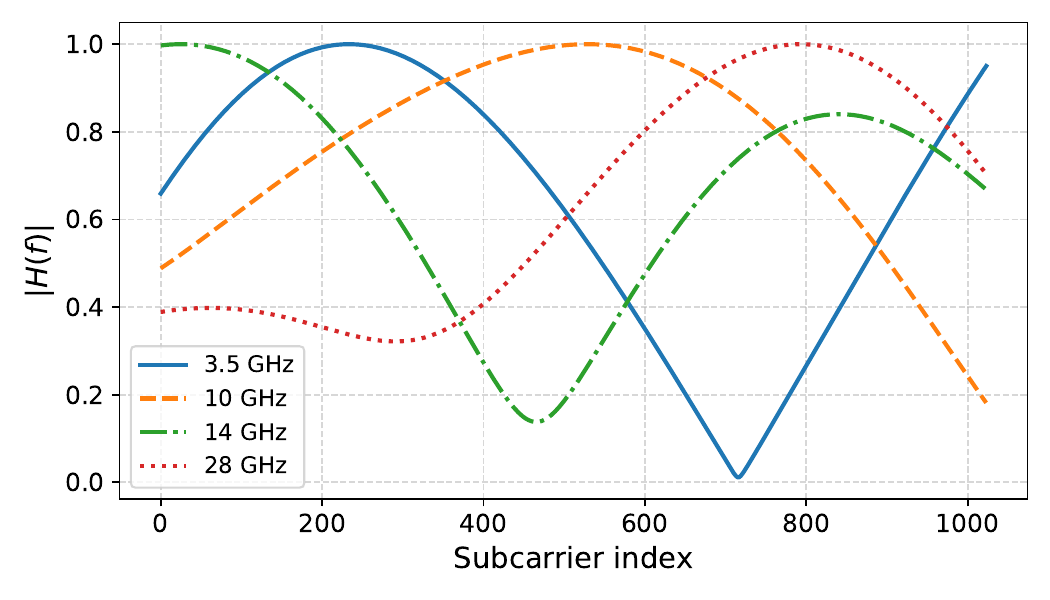}}
\caption{CFR examples for (a) indoor and (b) outdoor environments.}
\label{cfr}
\end{figure}

 \begin{figure}[!t]
  \centering
  {\includegraphics[width=3.5in]{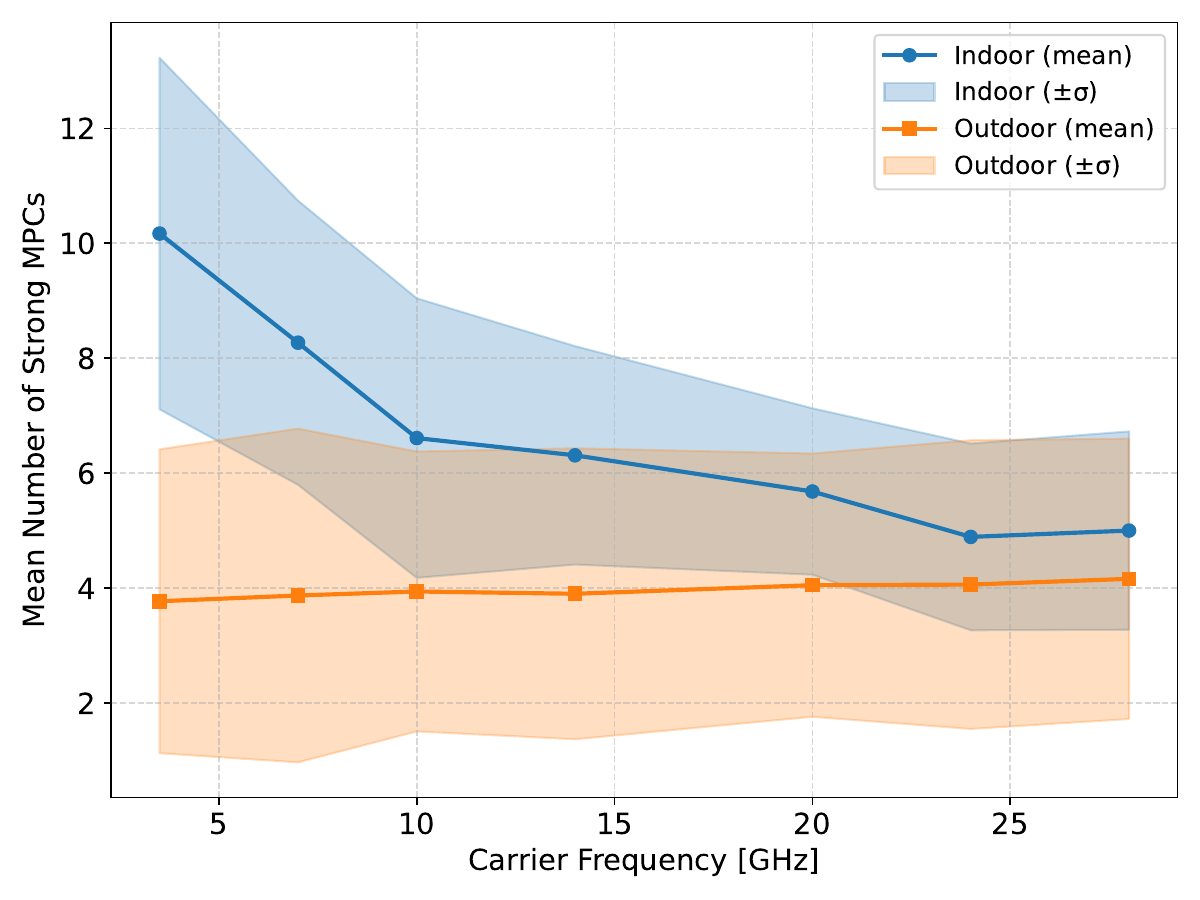}}
  \caption{MPC numbers versus frequency bands at two sites.} 
  \label{mpcnumber}
 \end{figure}

\begin{figure*}[!h]
\centering
\subfigure[][Indoor]{
   \includegraphics[width=2.3in]{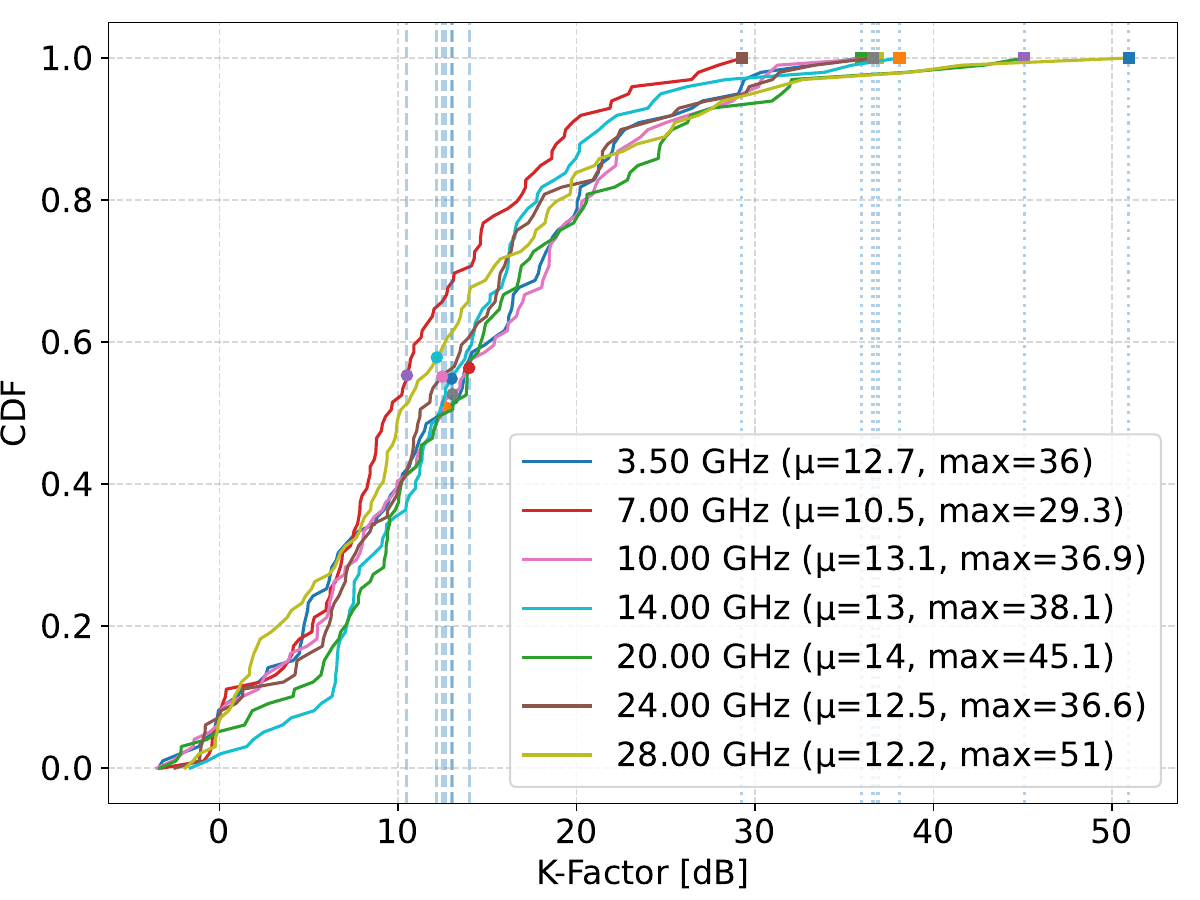}\label{kf-indoor}}
 \subfigure[][Indoor]{
    \includegraphics[width=2.3in]{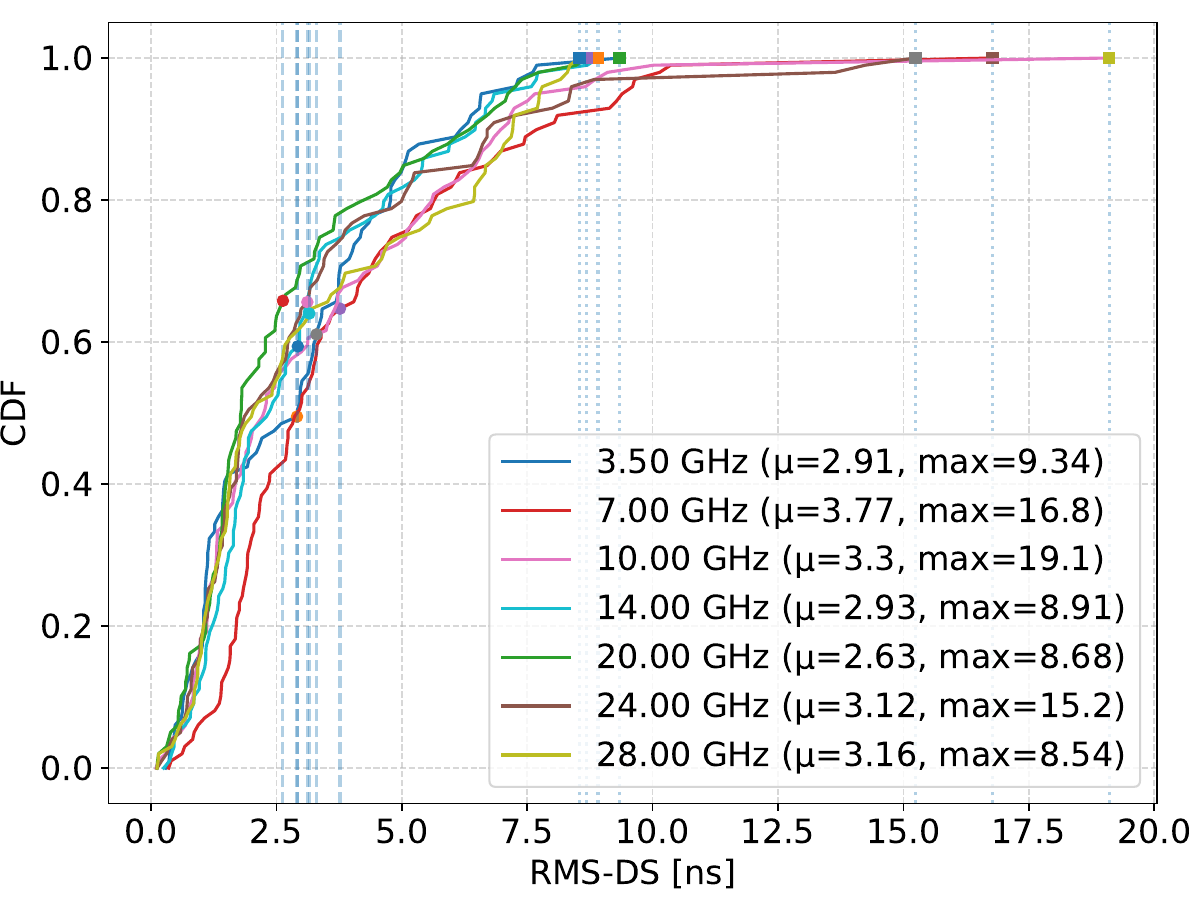}\label{RMSDS-indoor}}
\subfigure[][Indoor]{
   \includegraphics[width=2.3in]{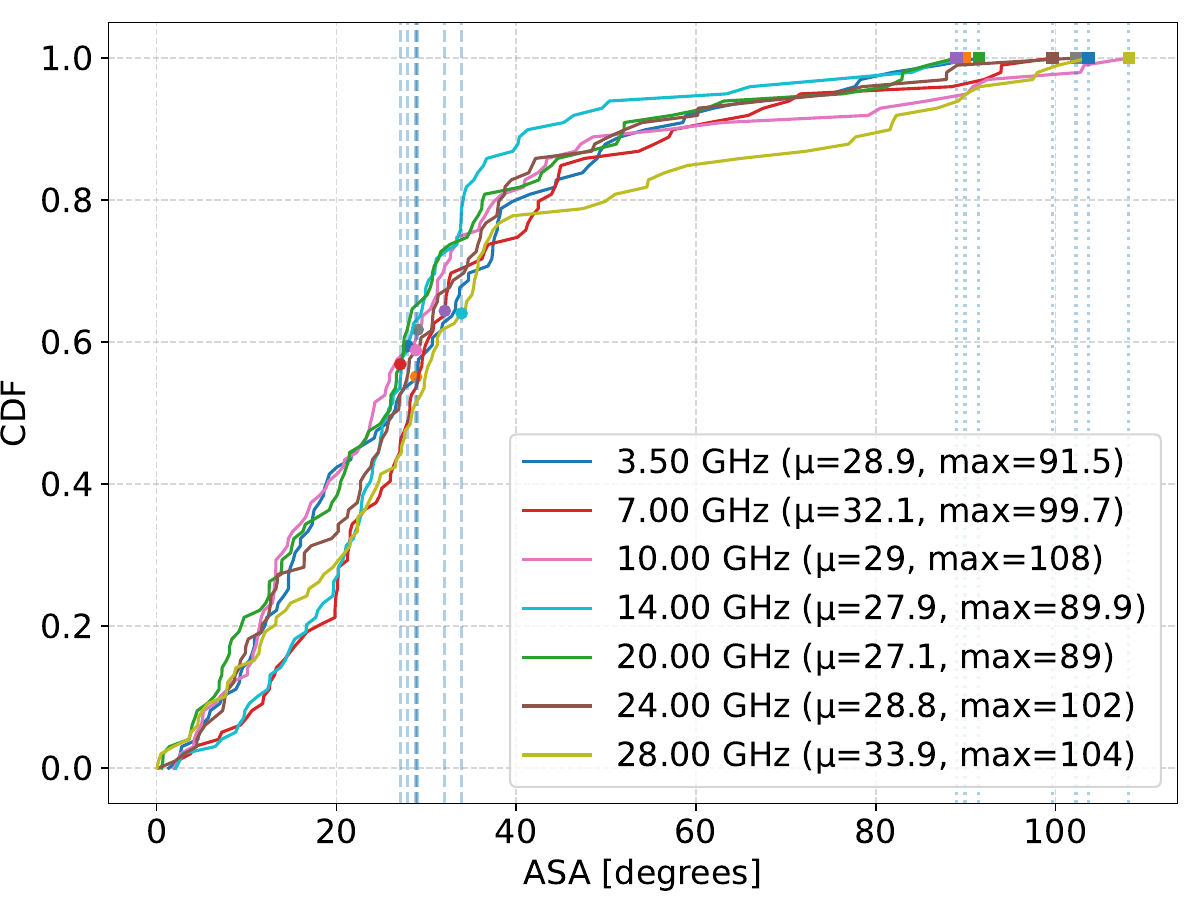}\label{asa-indoor}}
\subfigure[][Outdoor]{
   \includegraphics[width=2.3in]{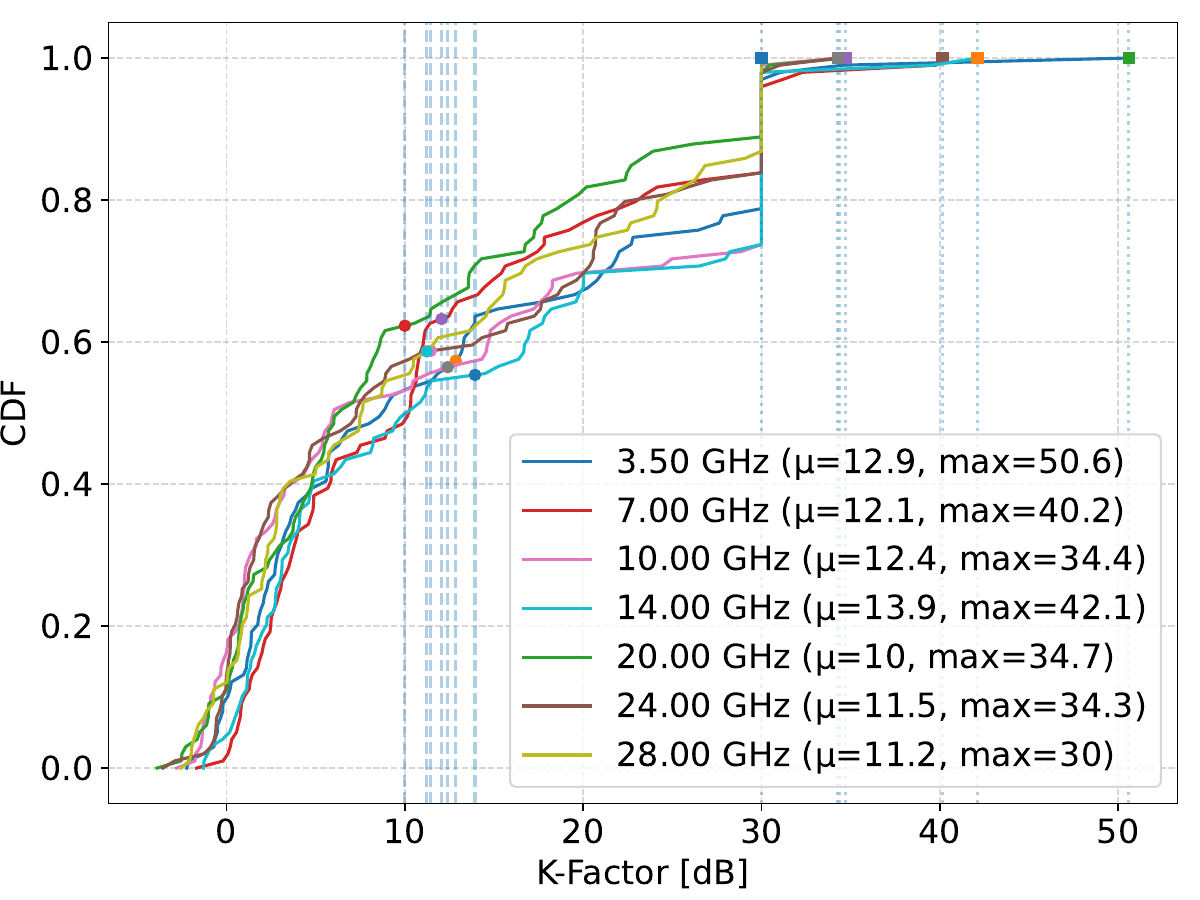}\label{kf-outdoor}}
\subfigure[][Outdoor]{
   \includegraphics[width=2.3in]{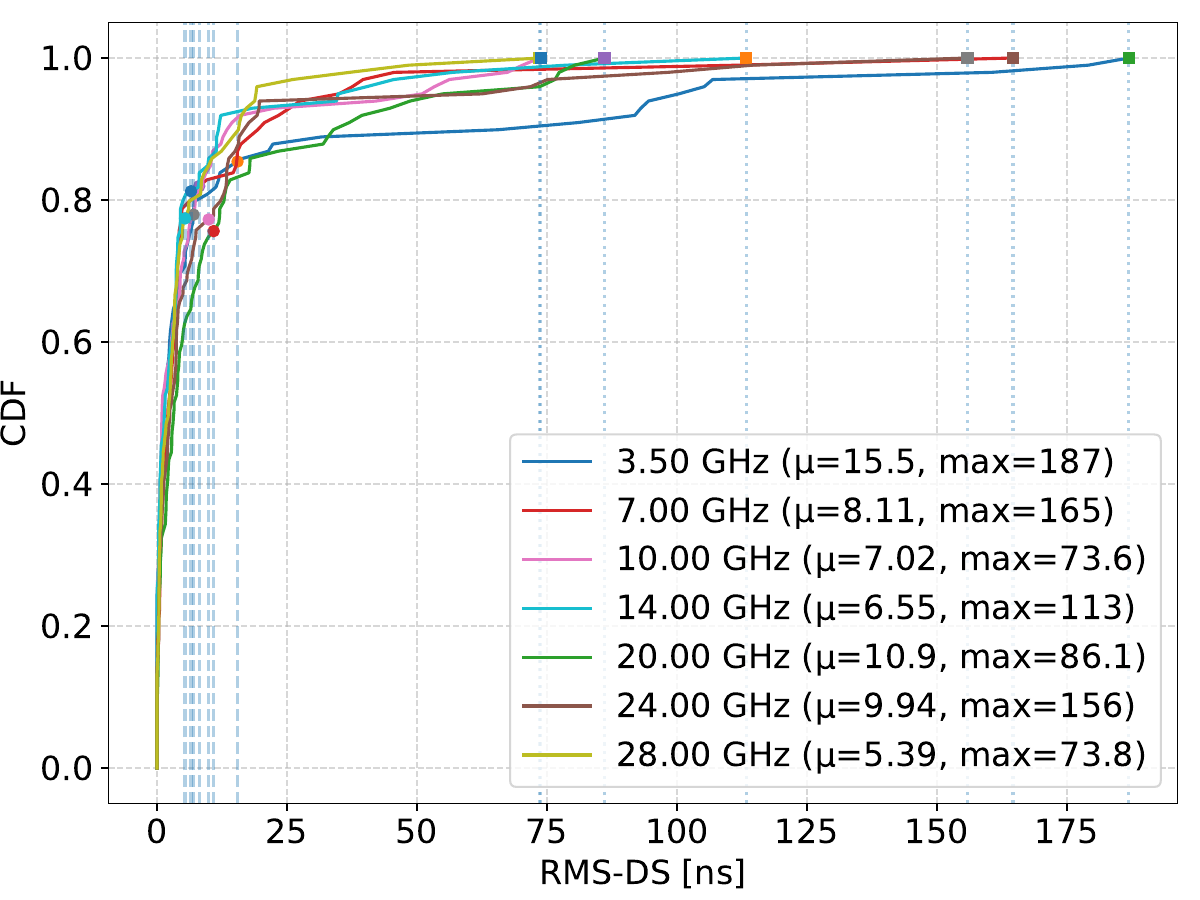}\label{RMSDS-outdoor}}
\subfigure[][Outdoor]{
   \includegraphics[width=2.3in]{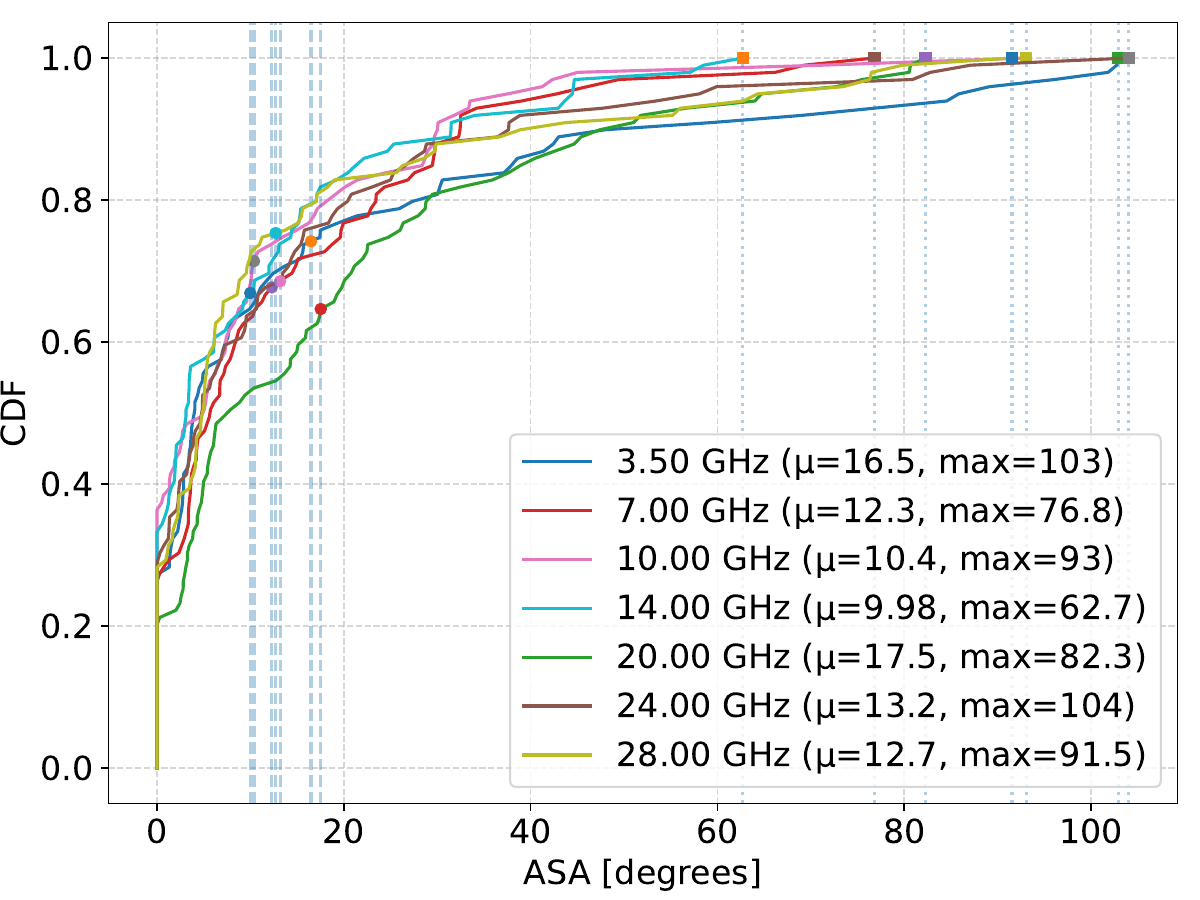}\label{asa-outdoor}}
\caption{CDF of channel characteristics including Rician K-factor, RMS-DS, and ASA.}
\label{channel_values}
\end{figure*}

\section{Channel Characterization}

In this section, we aim to reveal both frequency-dependent and site-specific behaviors of multipath propagation in both environments. We start from the power delay profiles (PDPs), and then focus on key channel metrics, including the Rician K-factor, delay-domain and angular-domain characteristics, which are mainly illustrated based on rays in CIR.

\subsection{PDPs}
The multipath structure of the channel is examined through the PDPs, which provide the temporal dispersion and relative strength of multipath components (MPCs). We consider a single antenna for both Tx and Rx, with the coordinates 
\begin{itemize}
    \item Indoor: Tx [4.8, 2.4, 1.0]~m, Rx [6.4, 6.8, 2.5]~m;
    \item Outdoor: Tx [-14, -6.0, 30.0]~m, Rx [15.0, 15.0, 1.5]~m. 
\end{itemize}
It is noted that PDP varies for different Tx or Rx locations, thus PDPs in Fig.~\ref{pdp} are obtained for the specific locations used above. However, there is a general trend that a higher number of indoor MPCs is observed than in the outdoor environment, confirming that confined spaces give rise to more complex propagation conditions. Furthermore, when considering the same bandwidth, i.e., 102.4 MHz, Fig.~\ref{cfr} shows that the indoor channels are more frequency selective than outdoor channels due to more MPCs. Regarding the frequency-dependent characteristics, Fig.~\ref{pdp} does not reveal a strong dependence on the frequency for the selected example locations. However, after averaging over 100 simulations with randomized Tx–Rx locations, Fig.~\ref{mpcnumber} shows that the number of MPCs decreases as a function of increasing frequencies in the indoor environment, while it remains mostly unchanged for outdoor scenarios. This behavior can be attributed to the limited dynamic range, which is set to 40~dB for both scenarios. In indoor environments, the relatively short Tx–Rx distances allow many weak MPCs to be captured, as shown in Fig.~\ref{pdp}(a)(b). As the frequency increases, the power of these weaker components diminishes, causing fewer MPCs to exceed the detection threshold. In contrast, outdoor environments are characterized by a much larger power disparity among MPCs, such that only a small number of dominant paths are retained, as shown in Fig.~\ref{pdp}(c)(d). As a result, increasing the frequency leads to only marginal changes in the observed number of MPCs in outdoor scenarios.

\subsection{Rician K-Factor}
We then consider a multi-antenna setup, using a $2\times2$ UPA for both Tx and Rx. We randomize Rx locations 100 times within environments, while Tx is fixed in the location used in Section III.A. We first focus on the Rician K-factor, characterizing the power ratio between the LoS and the NLoS components, providing insight into channel sparsity and LoS dominance. Fig.~\ref{kf-indoor} and Fig.~\ref{kf-outdoor} show the cumulative distribution functions (CDFs) of the K-factor for the indoor and outdoor scenarios, respectively, evaluated across different frequencies using a constant bandwidth of 102.4~MHz.

In the indoor scenario, the mean K-factor remains within a relatively narrow range, approximately between 10.5~dB and 14~dB across all considered frequencies. This indicates that increasing the frequency does not fundamentally alter the degree of LoS dominance in confined environments with short propagation distances. Nevertheless, higher frequencies, particularly those in the FR3 and FR2 bands, exhibit slightly higher upper-tail values. This suggests stronger LoS dominance in certain locations, which can be attributed to the reduction of NLoS components at higher frequencies, as shown in Fig.~\ref{mpcnumber}.

In the outdoor scenario, the K-factor distributions exhibit clear frequency-dependent behavior, as shown in Fig.~\ref{kf-outdoor}. While the mean K-factor values remain relatively stable across all considered frequencies (10--13.9~dB), the upper tail of the distribution changes noticeably. At lower frequencies, a small fraction of realizations exhibits very high K-factor values, indicating strongly LoS-dominated propagation conditions. As the carrier frequency increases, these extreme LoS-dominated cases become less prevalent, resulting in a reduced maximum K-factor and a more concentrated distribution. 

This behavior suggests that, although the number of dominant MPCs remains largely unchanged across frequencies in the outdoor environment, their relative power contributions evolve with frequency. Since the LoS component is normalized during processing, a lower observed K-factor implies an increase in the aggregate power of the NLoS components relative to the LoS. As a result, the FR3 bands retain moderate multipath contributions compared to the FR2 band, leading to comparable median K-factor values while exhibiting fewer highly LoS-dominated realizations. This power redistribution among MPCs is favorable for MIMO deployments in the FR3 band, as it supports spatial diversity and mitigates excessive channel dominance by a single propagation path.

\subsection{RMS Delay Spread}
The RMS-DS quantifies the temporal dispersion of the channel and directly affects inter-symbol interference. Fig.~\ref{RMSDS-indoor} and Fig.~\ref{RMSDS-outdoor} show the CDFs of the RMS-DS for the indoor and outdoor environments, respectively.

In the indoor scenario, the RMS-DS remains relatively small across all considered frequencies, with average values on the order of a few nanoseconds. Although some FR3 frequencies (e.g., 7~GHz, 10~GHz, and 24~GHz) exhibit larger maximum RMS-DS values due to occasional long reflection paths, the overall distributions are tightly clustered. This behavior reflects the confined geometry of indoor environments, where propagation distances are inherently bounded, and the RMS-DS is primarily governed by room dimensions rather than frequency.

In contrast, the outdoor scenario exhibits significantly larger RMS-DS values, particularly at lower frequencies. At 3.5~GHz, the mean RMS-DS exceeds 15~ns, with extreme cases extending beyond 180~ns. As the frequency increases, the RMS-DS generally decreases, driven by reduced diffraction efficiency and increased attenuation of long-delay MPCs.
Interestingly, a slight increase is observed around 20~GHz. This non-monotonic behavior can be attributed to the interplay between diminishing diffraction and relatively enhanced contributions from diffuse scattering. This indicates that the surface roughness of urban structures becomes comparable to the wavelength around 20~GHz, leading to increased angularly distributed scattered energy with moderate excess delays. At higher frequencies, such as 28~GHz, both diffraction and diffuse scattering are strongly attenuated, resulting in more compact delay profiles. 

Overall, these results indicate that the FR3 band offers a favorable trade-off: it substantially reduces delay dispersion compared to FR1, thereby easing wideband system design, while avoiding the extremely sparse delay profiles often observed in millimeter-wave bands.

\subsection{Azimuth Angular Spread of Arrival}

The azimuth angular spread of arrival (ASA) describes the dispersion of MPCs in the azimuth plane at the Rx and plays a crucial role in determining spatial multiplexing capability. Fig.~\ref{asa-indoor} and Fig.~\ref{asa-outdoor} illustrate the CDFs for the indoor and outdoor scenarios across different carrier frequencies.

In the indoor environment, the ASA remains relatively large across all frequency bands, with median values typically in the range of 28--33$^\circ$. This indicates rich angular diversity caused by dense scattering from walls, furniture, and other objects. Notably, the ASA does not exhibit a strictly monotonic dependence on frequency, highlighting that indoor angular dispersion is primarily governed by the room geometry, similar to the behavior observed for RMS-DS.

In the outdoor scenario, the ASA exhibits a clearer frequency dependency, consistent with measurements in \cite{fr3_mea_outdoor}. At lower frequencies, strong diffraction around buildings lead to wide angular spreads. As the frequency increases toward the \textit{lower FR3 range}, diffraction effects gradually weaken, reducing the contribution of long-range diffracted paths. At the same time, diffuse scattering from building facades and street-level structures becomes relatively more pronounced due to increased surface roughness relative to the wavelength, resulting in a temporary increase in angular dispersion. This explains the observed rise in ASA around 20~GHz.

Beyond this frequency, propagation becomes increasingly dominated by a small number of highly directional specular paths, while both diffraction and diffuse scattering are strongly attenuated. Consequently, the angular distribution of the channel becomes more concentrated, and the ASA decreases again as the frequency approaches 28~GHz.

The moderate angular spread in FR3 is particularly beneficial for MIMO systems, as it enables effective spatial multiplexing while allowing for high-resolution beamforming. This angular-domain behavior further explains the favorable MIMO performance of FR3 observed in subsequent sections.

\subsection{Discussion: Frequency Dependency and Site Specificity}

The above channel characterization results highlight the joint impact of carrier frequency and site geometry on multipath propagation. Across all considered metrics, channel behavior exhibits clear frequency dependency while remaining strongly influenced by the propagation environment. 

From a frequency perspective, increasing the frequency generally reduces diffraction efficiency and attenuates long-delay and wide-angle MPCs. This leads to higher relative LoS dominance, more compact delay profiles, and narrower angular spreads at higher frequencies. However, the FR3 band does not follow a purely monotonic evolution from FR1 to FR2. Instead, intermediate frequencies exhibit distinct propagation characteristics, such as the local increase in RMS-DS and ASA around 20~GHz.

From a site perspective, the indoor and outdoor scenarios exhibit fundamentally different propagation behaviors. In indoor environments, confined geometry and dense multipath dominate channel characteristics, resulting in relatively stable channel characteristics across frequencies. In contrast, outdoor urban environments are highly site-specific, where building layout, street orientation, and user location strongly influence the relative contributions of LoS, reflected, and scattered paths, leading to larger variations across frequency bands.

Overall, FR3 propagation is governed by a complex interplay among reflection, diffraction, and scattering mechanisms, rather than a simple transition from rich to sparse multipath conditions. The FR3 band strikes a favorable balance between multipath richness and propagation compactness, making it particularly attractive for MIMO systems.

\section{MIMO Performance}

In this section, we evaluate the MIMO performance across different frequencies based on the CFRs. By leveraging the site-specific channel data, we examine how frequency, array configuration, and propagation jointly affect spatial multiplexing and channel capacity. Unless otherwise stated, the results are averaged over subcarriers within the system bandwidth.

\subsection{Performance Metrics}

We employ three widely used performance metrics: channel rank, condition number, and SE. Together, these metrics characterize the spatial multiplexing capability, channel conditioning, and achievable throughput of MIMO systems.

\subsubsection{Channel Rank}
The channel rank reflects the maximum number of parallel data streams that can be supported by the propagation channel and antenna configuration. It captures the richness of the spatial channel and is directly influenced by the angular and delay dispersion of MPCs. For a given carrier frequency $f_c$, user location $p$, and OFDM subcarrier $k$, the MIMO channel matrix is denoted by $\mathbf{H}_{p,k}(f_c) \in \mathbb{C}^{N_r \times N_t}$. Its singular value decomposition (SVD) is given by
\begin{equation}
\mathbf{H}_{p,k}(f_c)
= \mathbf{U}_{p,k}\,\mathbf{\Sigma}_{p,k}\,\mathbf{V}_{p,k}^H,
\end{equation}
where $\mathbf{\Sigma}_{p,k} =\mathrm{diag}\!\left(\sigma_{p,k,1}, \ldots, \sigma_{p,k,m}\right)$,
with $m = \min(N_r, N_t)$
and the singular values satisfy
$\sigma_{p,k,1} \ge \sigma_{p,k,2} \ge \cdots \ge \sigma_{p,k,m} \ge 0$. The \emph{effective MIMO rank} is defined using a relative threshold $\zeta \in (0,1)$ as
\begin{equation}
r_{p,k}(f_c;\zeta)
= \sum_{i=1}^{m}
\mathbf{1}
\left\{
\sigma_{p,k,i}
\ge
\zeta \, \sigma_{p,k,1}
\right\},
\end{equation}
where $\mathbf{1}\{\cdot\}$ denotes the indicator function. For each carrier frequency, the channel rank is obtained by averaging $r_{p,k}$ over all subcarriers and user locations, using $\zeta = 10^{-3}$.

\subsubsection{Condition Number}
The condition number of the MIMO channel matrix is defined as the ratio between the largest and smallest singular values and serves as an indicator of spatial channel orthogonality and robustness to noise and estimation errors. A smaller condition number implies better-conditioned channels and improved performance for spatial multiplexing and linear precoding schemes. This metric is particularly useful for assessing spatial correlation effects. The condition number is defined as
\begin{equation}
\kappa_{p,k}(f_c) =
\frac{\sigma_{p,k,1}}{\sigma_{p,k,m}},
\end{equation}
where $\sigma_{p,k,1}$ and $\sigma_{p,k,m}$ denote the largest and smallest singular values of $\mathbf{H}_{p,k}(f_c)$, respectively.

\subsubsection{Spectral Efficiency}
SE quantifies the achievable rate per unit bandwidth and provides an overall measure of MIMO performance. For the $p$-th user location, it is computed as
\begin{equation}
\begin{aligned}
C_p(f_c)
&= \frac{1}{N_f} \sum_{k=1}^{N_f}
\log_2 \det \!\left(
\mathbf{I}
+ \frac{\rho}{N_t}
\mathbf{H}_{p,k}(f_c)\mathbf{H}_{p,k}^{\mathrm{H}}(f_c)
\right) \\
&= \frac{1}{N_f}
\sum_{k=1}^{N_f}
\sum_{i=1}^{r_{p,k}}
\log_2 \!\left(
1 + \frac{\rho}{N_t}\sigma_{p,k,i}^2
\right),
\end{aligned}
\end{equation}
where $N_f$ is the number of subcarriers, $N_t$ is the number of transmit antennas, and $\rho$ denotes the average signal-to-noise ratio (SNR), which is set as 20~dB for the evaluation.

These three metrics are complementary: while the channel rank captures the maximum spatial degrees of freedom (DoFs), the condition number reflects the quality and orthogonality of these DoFs, and the SE integrates both aspects into an end-to-end performance measure. Together, they provide a comprehensive evaluation of MIMO performance across FR1, FR3, and FR2 bands.

 \begin{figure*}[!t]
  \centering
   \subfigure[Indoor scenario: Lab room]{\includegraphics[width=0.92\linewidth]{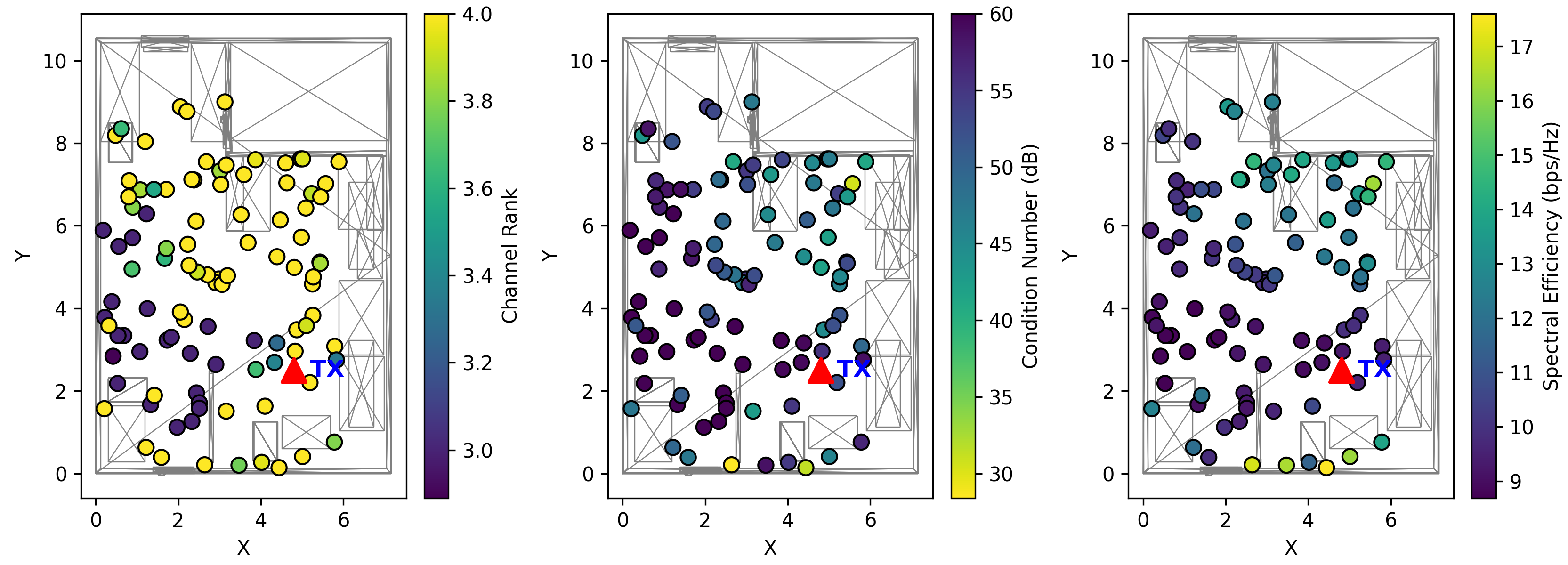}\label{fig:site_indoor}}
   \subfigure[Outdoor scenario: Urban city]{\includegraphics[width=0.92\linewidth]{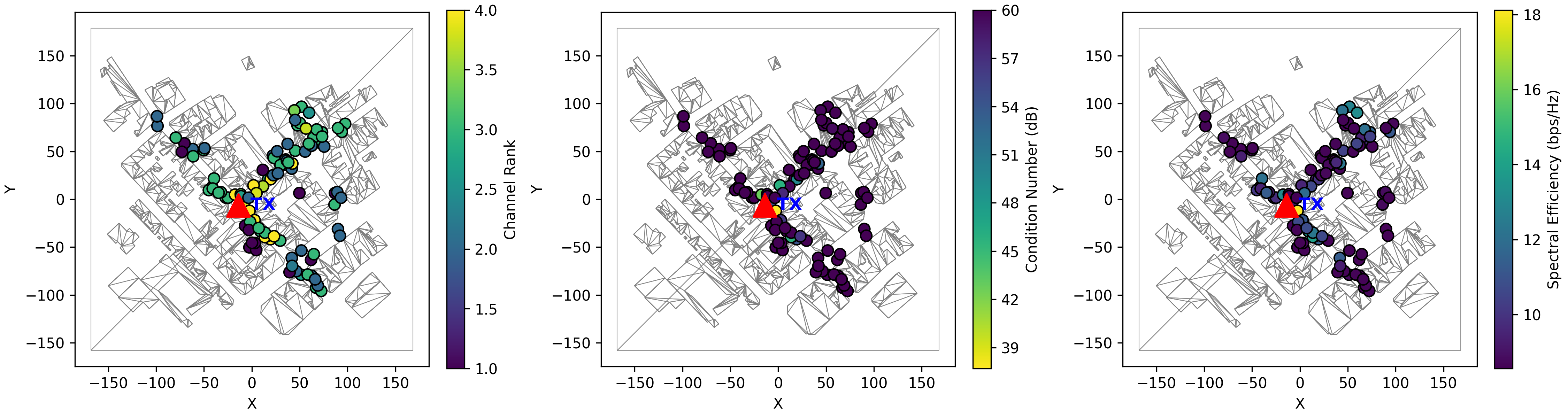}\label{fig:site_outdoor}}
  \caption{Site-location-specific visualization of channel rank, condition number, and SE in the (a) indoor and (b) outdoor scenarios at 10~GHz.}
  \label{performance_overview}
 \end{figure*}
\subsection{Site-Location-specific Performance Visualization}
To further illustrate the site-specific nature of MIMO performance, we visualize key performance metrics spatially over the considered indoor and outdoor environments. Fig.~\ref{fig:site_indoor} and Fig.~\ref{fig:site_outdoor} show the spatial distributions of channel rank, condition number, and SE at 10~GHz for the outdoor and indoor scenarios, respectively. Each point corresponds to an Rx location, while the Tx is fixed and indicated by the red triangle. The system bandwidth is set to 102.4~MHz, and a $2\times2$ UPA is employed at both the Tx and Rx. For visualization purposes, the condition number is capped at 60~dB.

In the indoor scenario, the spatial variations are noticeably smoother compared to the outdoor case. Owing to the confined geometry and dense scattering from walls and furniture, most locations experience relatively high channel ranks ($\geq 2.5$), while the rank can be as low as 1 in the outdoor scenario. As a result, the achievable SE remains more uniform across the room, with fewer extreme performance degradations caused by poor condition numbers. This behavior indicates that indoor environments are less sensitive to the receiver location because of the confined geometry. 

In the outdoor scenario, the performance metrics exhibit strong spatial variations that are closely tied to the surrounding geometry. Locations with rich multipath propagation, typically caused by multiple building reflections and favorable street orientations, achieve higher channel rank and SE, together with lower condition numbers. In contrast, Rx locations dominated by a LoS or a limited number of MPCs show reduced channel rank and poorly conditioned channels, leading to degraded SE. This pronounced spatial heterogeneity highlights the strong location specificity of outdoor FR3 channels, where local building layout and street orientation play a dominant role.

Overall, these site-specific visualizations demonstrate that MIMO performance at FR3 is highly dependent on the environment. While indoor scenarios benefit from stable and spatially consistent performance, outdoor deployments require careful consideration of site geometry and user location. These observations motivate the need for adaptive MIMO architectures, particularly for outdoor FR3 systems.

\begin{figure*}[!t]
\centering
\subfigure[][]{
   \includegraphics[width=2.3in]{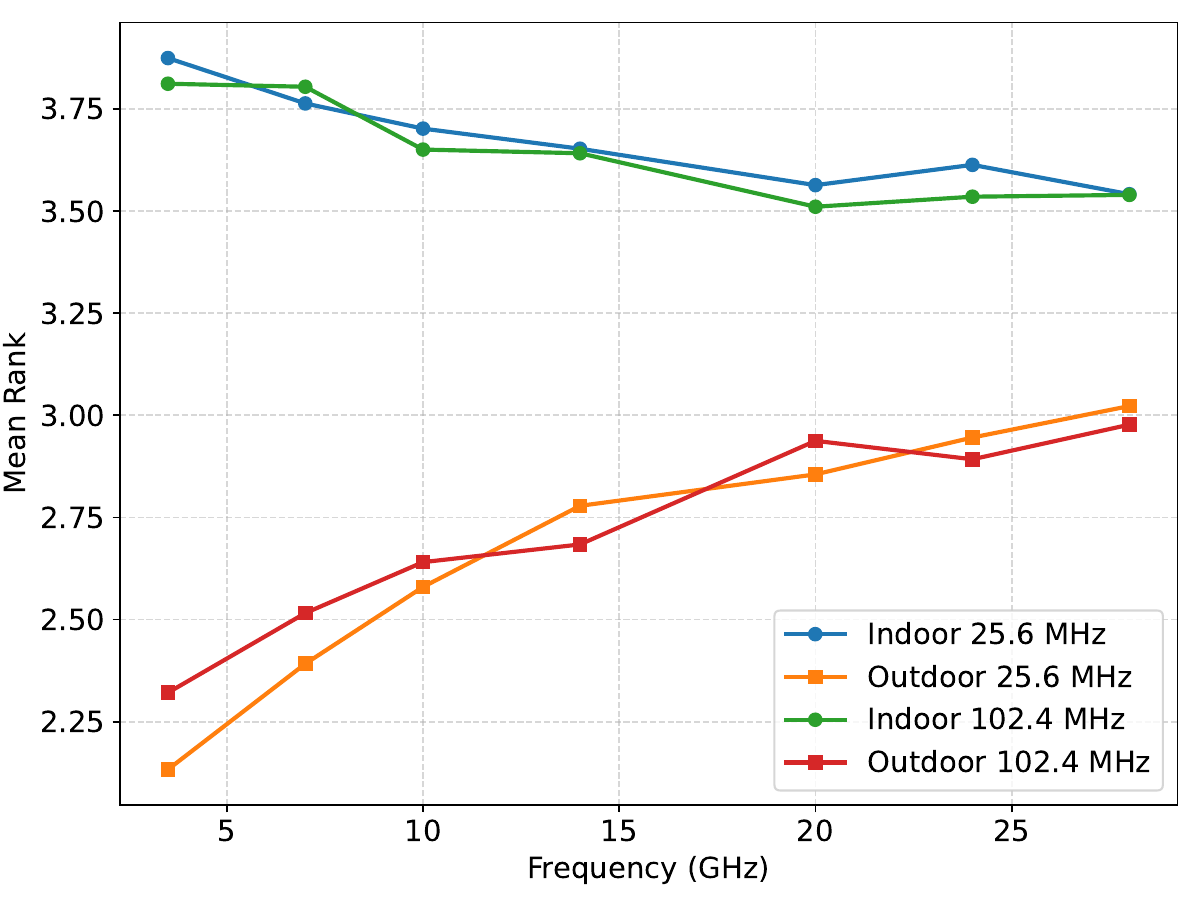}\label{meanrank}}
 \subfigure[][]{
    \includegraphics[width=2.3in]{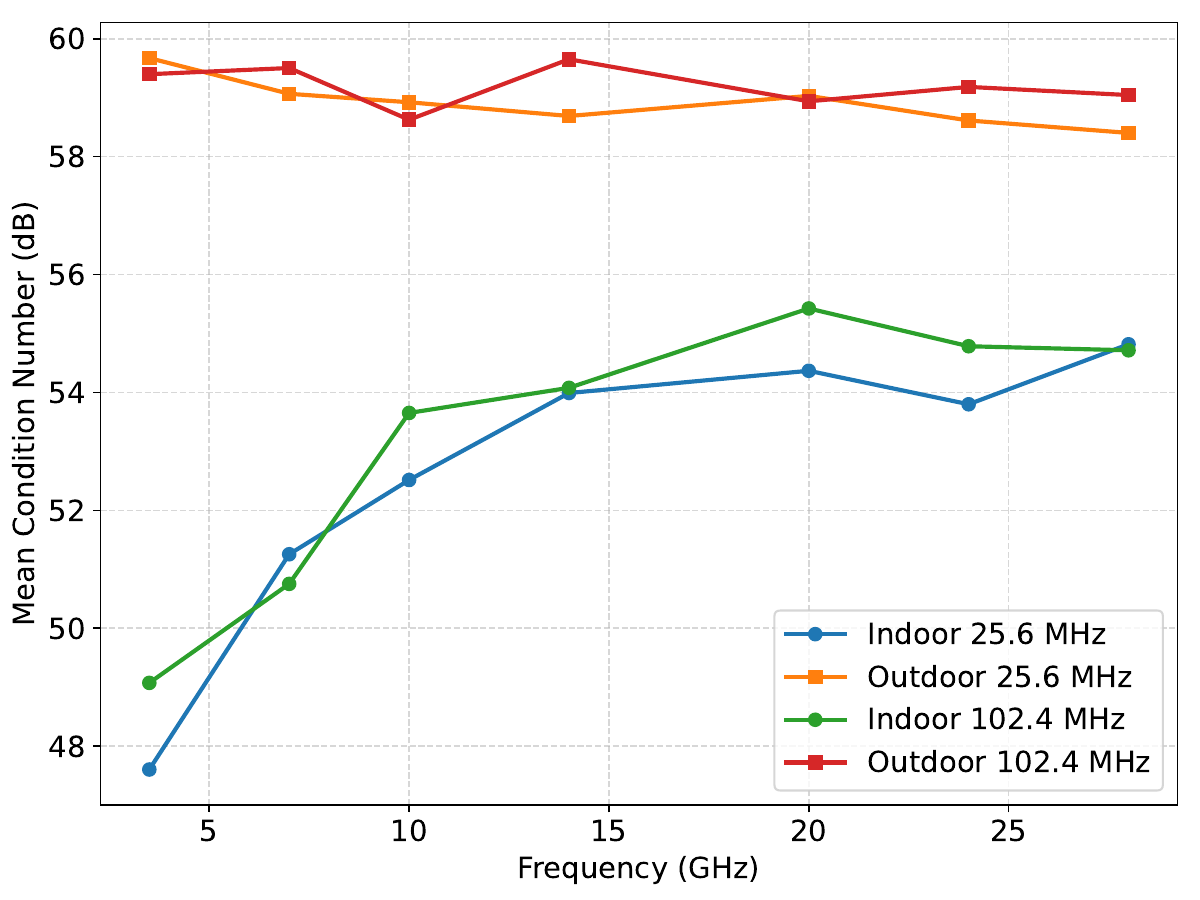}\label{meancn}}
\subfigure[][]{
   \includegraphics[width=2.3in]{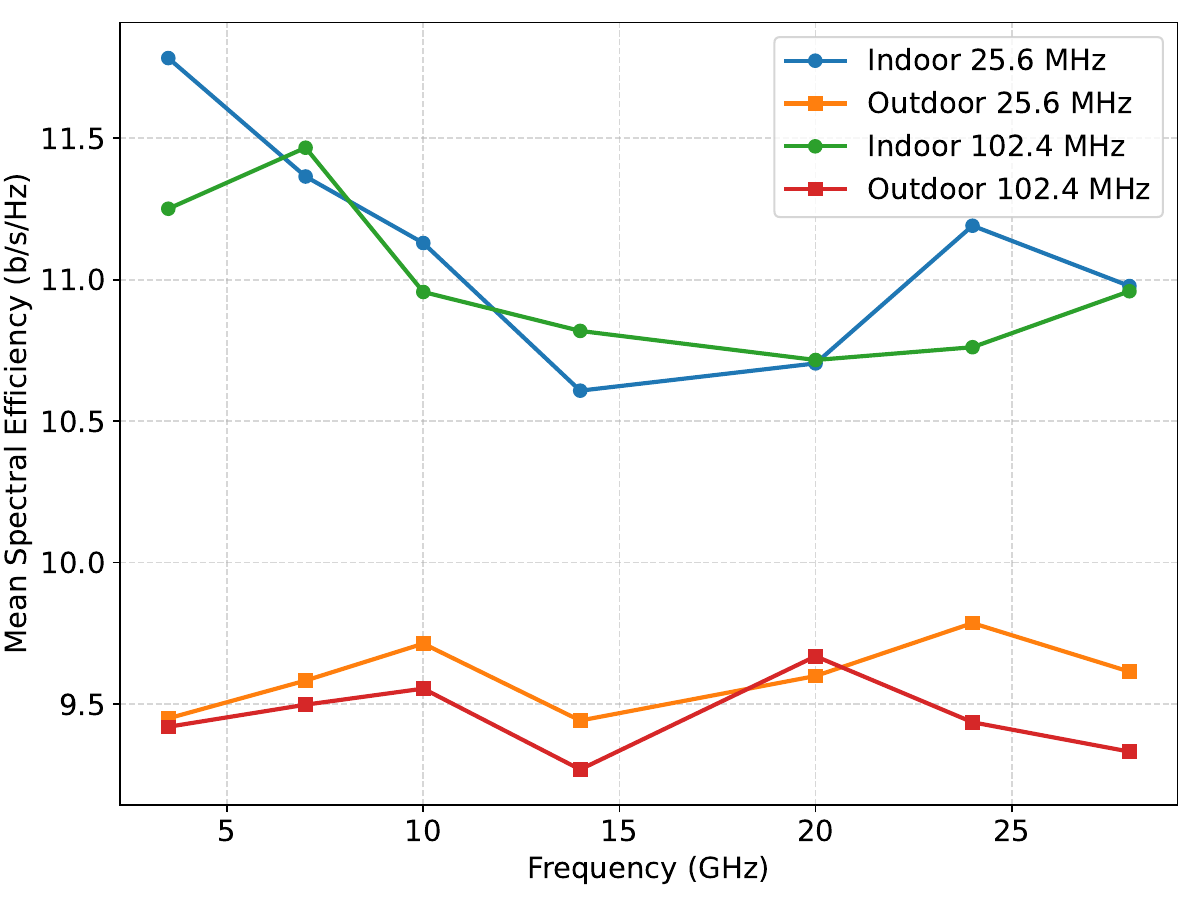}\label{meanse}}
\caption{Frequency-dependent MIMO performance: a) channel rank, b) condition number, and c) SE.}
\label{mimoperformance}
\end{figure*}

\subsection{Frequency-Dependent Performance}

We next examine how MIMO performance evolves with carrier frequency by evaluating the mean channel rank, mean condition number, and mean SE across the considered bands. Fig.~\ref{mimoperformance} summarizes the results for both indoor and outdoor scenarios under two practical bandwidth configurations.

\subsubsection{Channel Rank}
As shown in Fig.~\ref{meanrank}, the mean channel rank exhibits a clear frequency dependence in both environments. In the indoor scenario, the channel rank remains relatively high across the entire frequency range, with only a mild decreasing trend as the carrier frequency increases. This behavior is consistent with the confined geometry and dense multipath in indoor environments, which preserve spatial DoFs even at higher frequencies.

In contrast, the outdoor scenario exhibits an increasing trend in channel rank with carrier frequency. 
Although the number of dominant MPCs in the outdoor environment remains approximately constant across frequencies, their spatial resolvability improves at higher frequencies. As the wavelength decreases, phase variations across the antenna elements become more sensitive to angular differences between incoming paths, leading to an increase in the effective channel rank. 

Consequently, the FR3 bands achieve higher channel ranks than the FR1 band and approach or exceed the performance observed at FR2, demonstrating their capability to support spatial multiplexing in outdoor deployments.

\subsubsection{Condition Number} Since we aim to show the trend of ranks and condition numbers, we employ a small threshold ($\zeta = 10^{-3}$) to have enough changes. Although it leads to a high condition number, the observed evolving trends are more valuable. 
Fig.~\ref{meancn} reveals a strong site dependency. In the outdoor environment, the mean condition number remains consistently high (approximately 58--59~dB) across all considered frequencies, indicating strongly correlated spatial subchannels. This behavior persists despite the slight increase in channel rank observed at higher frequencies. Although the number of dominant MPCs in the outdoor scenario remains approximately constant, their spatial resolvability improves modestly at higher frequencies due to increased angular separation and reduced diffraction effects. However, under the considered antenna configuration, i.e., $2 \times 2$ UPA with half-wavelength spacing, the array lacks sufficient spatial resolution to fully exploit the improved angular separability of the MPCs. Consequently, the variance of singular values remains pronounced, and the channel remains poorly conditioned.

In the indoor scenario, the mean condition number increases gradually from FR1 to FR3. This trend is consistent with the mild reduction in channel rank and reflects increased spatial correlation as the effective multipath richness diminishes at higher frequencies in a confined environment.

Furthermore, increasing the bandwidth from 25.6~MHz to 102.4~MHz leads to a small but noticeable increase in the mean condition number. This indicates that wider bandwidths do not necessarily improve spatial conditioning under the fixed array geometry, and that spatial correlation remains a fundamental limiting factor in site-specific MIMO channels.

\subsubsection{Spectral Efficiency}
The frequency-dependent trends in SE are shown in Fig.~\ref{meanse}. In the indoor scenario, the SE exhibits a mild decreasing trend with increasing frequency. This behavior is primarily caused by the reduction in spatial multiplexing capability and the increase in channel correlation at higher frequencies, as observed from the channel rank and condition number results. Overall, it suggests that low-frequency bands (FR1) are still promising in the indoor environment. Nevertheless, the performance degradation remains moderate within the lower part of the FR3 band (below 10 GHz), and the achieved SE remains comparable to that of sub-6~GHz deployments under the considered array configuration. 

In the outdoor scenario, SE remains relatively stable across frequencies, reflecting the combined effects of increasing spatial multiplexing capability and high channel correlation. Importantly, the FR3 bands achieve slightly higher SE than both FR1 and FR2. These results confirm that FR3 offers a favorable performance trade-off across a wide range of deployment scenarios, e.g., outdoor scenarios.

Overall, the frequency-dependent analysis demonstrates that MIMO performance does not degrade monotonically with increasing frequency. Instead, FR3 occupies a transitional regime that preserves spatial multiplexing gains while enabling higher SE, making it a promising candidate band for future dynamic-spectrum and adaptive MIMO systems.

\section{XL-MIMO in FR3}
In this section, we investigate the potential of FR3 from conventional to XL-MIMO systems by examining how system performance scales with the antenna array size. We begin by analyzing channel hardening as a function of the number of antennas. Building on this fundamental insight, we then evaluate the scaling behavior of key MIMO performance metrics under two physically meaningful array-scaling laws: fixed antenna count and fixed physical aperture.

\subsection{Channel Hardening}
Channel hardening refers to the phenomenon whereby the impact of small-scale fading diminishes as the number of antennas increases, such that the effective channel gain becomes increasingly deterministic. To quantify this effect, we consider the equivalent channel gain after matched filtering. For a single-user MIMO link, the effective channel gain
on subcarrier $k$ is given by the squared Frobenius norm $\|\mathbf{H}_k\|_F^2$. As channel hardening occurs, $\|\mathbf{H}_k\|_F^2$ concentrates around its mean,
which allows the instantaneous SNR to be accurately approximated by its average value. Accordingly, the instantaneous SE on subcarrier $k$ can be expressed as $C_k=\log_2\!\left(1 + \rho \, \|\mathbf{H}_k\|_F^2 \right).$
Under strong channel hardening, the random channel gain $\|\mathbf{H}_k\|_F^2$
can be approximated by its expectation, yielding
\begin{equation}
C_k
\approx
\log_2\!\left(
1 + \rho \, \mathbb{E}\!\left[\|\mathbf{H}_k\|_F^2\right]
\right).
\label{eq:se_hardened}
\end{equation}
This approximation becomes increasingly accurate as the normalized variance of
$\|\mathbf{H}_k\|_F^2$ decreases with an increasing number of antennas ($\mathbf{H}_{k} \in \mathbb{C}^{N_r \times N_t}$).

For frequency-selective channels, the channel hardening metric is evaluated per
subcarrier and averaged over the system bandwidth. Specifically, the metric is defined as \cite{charden}
\begin{equation}
\eta_{\mathrm{wb}}
\triangleq
\frac{1}{N_f}
\sum_{k=1}^{N_f}
\frac{\mathrm{Var}\!\left(\|\mathbf{H}_k\|_F^2\right)}
{\left(\mathbb{E}\!\left[\|\mathbf{H}_k\|_F^2\right]\right)^2}.
\end{equation}

To illustrate the channel hardening phenomenon, we conducted a simulation for the outdoor urban scenario with fixed Tx and Rx locations (in LoS condition), but varying Tx antenna arrays from $2\times2$ to $15\times15$ while using $3\times3$ at Rx, both with a half-wavelength spacing of elements. For each configuration, the channel hardening metric $\eta$ is computed from the frequency-domain channel responses by evaluating the normalized variance of the channel gain across subcarriers, and is expressed in decibels as $10\log_{10}(\eta)$.

\begin{figure}[!t]
  \centering
  {\includegraphics[width=3.5in]{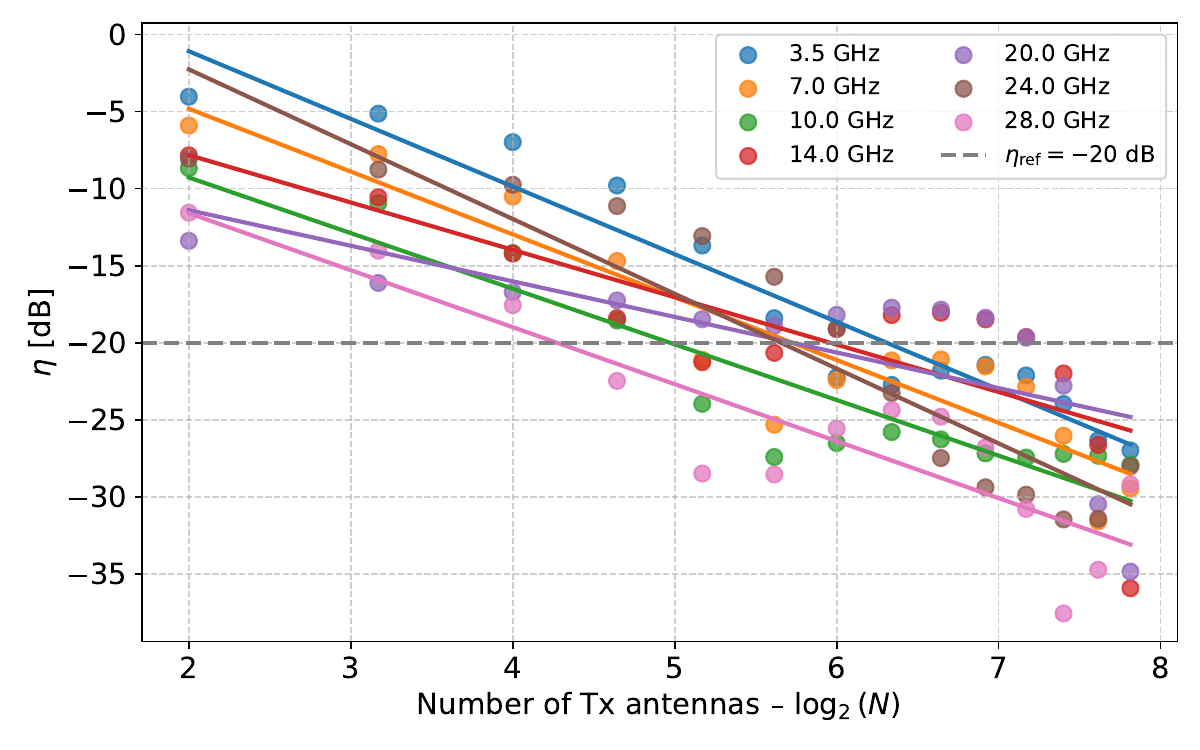}}
  \caption{$\eta$ versus the number of Tx antennas.} 
  \label{fig:channelharden}
 \end{figure}

As shown in Fig.~\ref{fig:channelharden}, the channel hardening metric $\eta$ decreases monotonically with increasing antenna count for all frequencies, indicating progressively stronger channel hardening as the array size grows. When plotted versus $\log_2(N)$, the results across all bands collapse onto the linear trends, 
consistent with reported measurements \cite{tang_hardening}. Using $-20$~dB as a reference threshold for effective channel hardening \cite{charden}, the least-squares fitted curves indicate that the same level of channel hardening can be achieved with different numbers of antennas across frequencies. Moreover, observations at different Rx locations reveal variations in the channel hardening trend, indicating that channel hardening is also influenced by location-specific propagation conditions, such as MPC's richness, rather than by carrier frequency alone.

 \begin{figure}[!t]
  \centering
  {\includegraphics[width=3.5in]{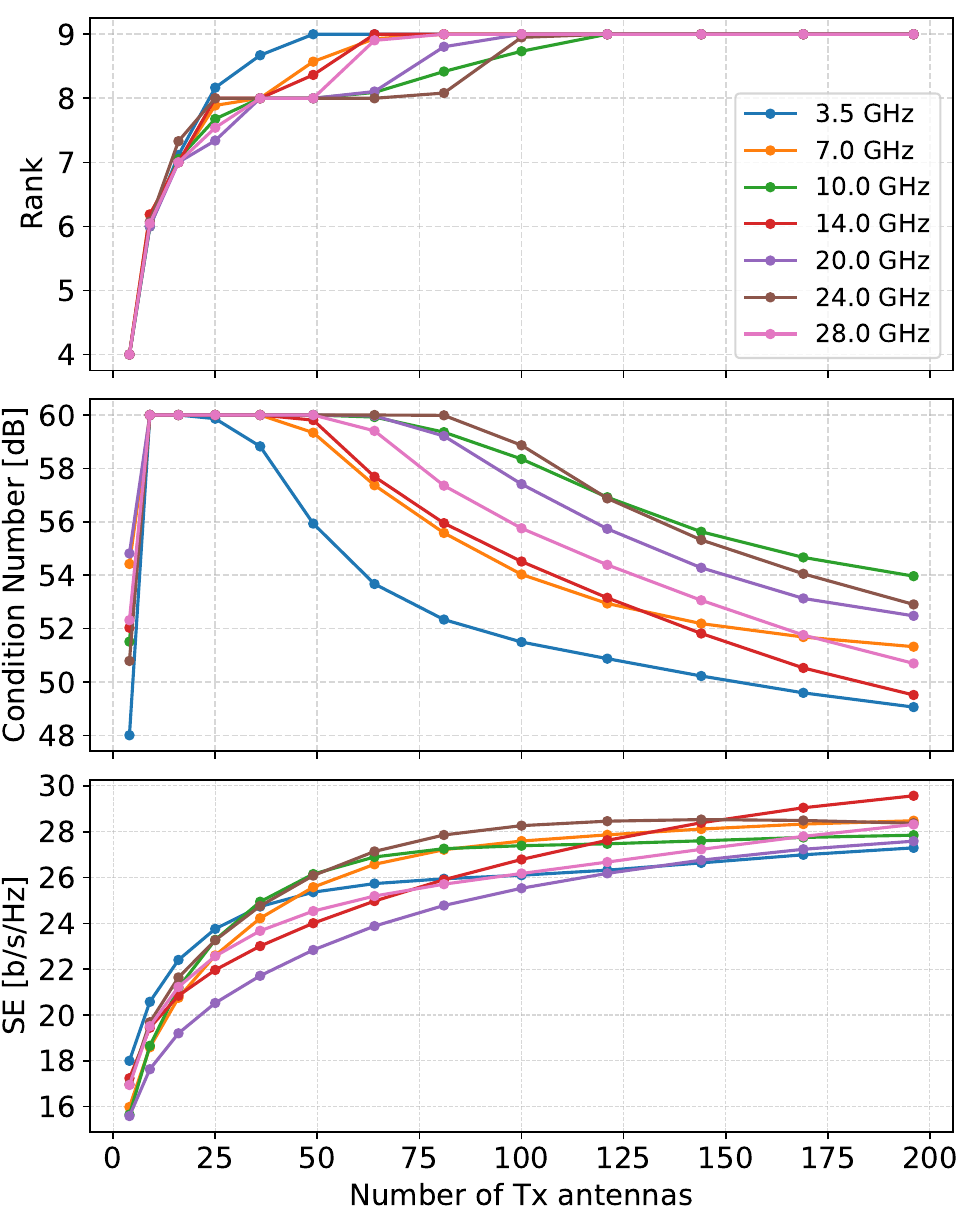}}
  \caption{XL-MIMO performance considering a fixed number of antennas over frequencies.} 
  \label{mimo_size}
 \end{figure}
\subsection{Performance with Fixed Number of Antennas}

Following on that, Fig.~\ref{mimo_size} illustrates the MIMO performance as a function of the number of Tx antennas, where the antenna count is fixed across all frequency bands. As the array size increases, the channel rank rapidly increases and then saturates for all frequencies.
As the Tx array grows, the condition number decreases monotonically, indicating improved channel orthogonality and more balanced singular values. This improvement is more pronounced at lower frequencies for the given size of array, where the physical aperture is larger at low frequencies, increasing angular resolution and reducing spatial correlation among antenna elements. 

The SE increases with the number of Tx antennas across all frequencies, but with diminishing returns as the array size becomes large. This saturation behavior is consistent with the simultaneous saturation of channel rank and the bounded receive-side spatial DoF. Among the considered bands, FR3 frequencies generally achieve SE values that are comparable to or higher than those of FR1 and FR2 at large array sizes (e.g., $N_t\geq 36)$, demonstrating a favorable balance between spatial multiplexing capability and channel conditioning.

\subsection{Performance with Fixed Physical Aperture}

\begin{figure}[!t]
  \centering
  {\includegraphics[width=3.4in]{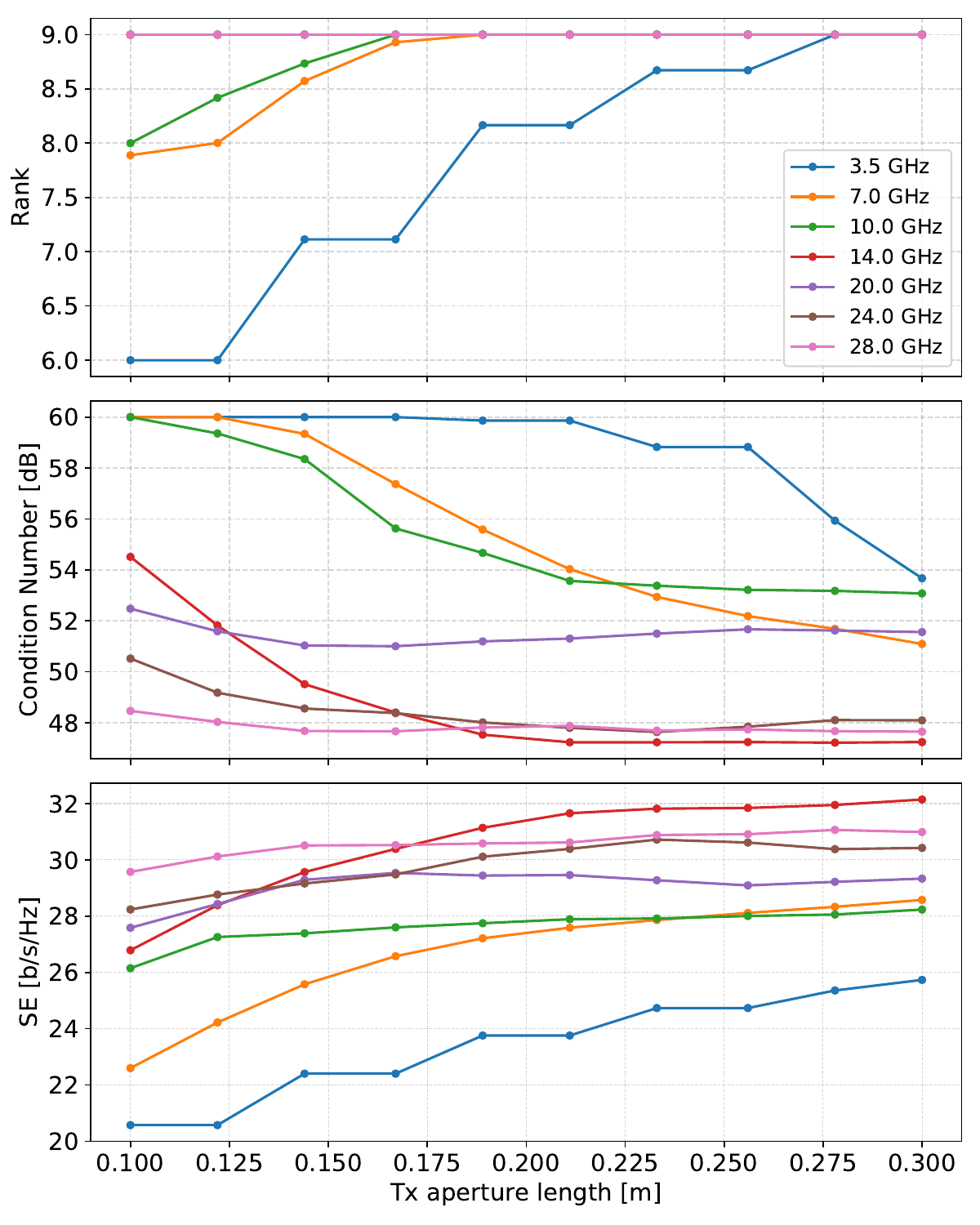}}
  \caption{XL-MIMO performance with a fixed physical array aperture over frequencies.} 
  \label{mimo_aperture}
 \end{figure}

A fundamentally different performance trend is observed when the physical aperture of the Tx array is fixed, as shown in Fig.~\ref{mimo_aperture}. Under this scaling law, the inter-element spacing is kept at half-wavelength, such that the number of antenna elements increases with frequency due to the shorter wavelength. Thus, higher-frequency bands can accommodate significantly denser arrays within the same physical aperture.

Under this fixed-aperture scaling, FR3 clearly benefits from the increased antenna density. The channel rank reaches its saturation value at smaller physical aperture sizes compared with FR1, indicating improved spatial resolvability of MPCs at higher frequencies. In addition, the condition number decreases as the aperture grows, particularly for FR3 and FR2, reflecting enhanced channel orthogonality and stronger channel hardening enabled by more antenna elements.

Most importantly, the SE exhibits a pronounced frequency advantage under this comparison. FR3 consistently outperforms FR1 and achieves performance comparable to, or even exceeding, that of FR2 across a wide range of aperture sizes. The result indicates that FR3 can effectively exploit the spatial DoFs offered by XL-MIMO arrays that are barely feasible in FR1, while avoiding unfavorable propagation in FR2.

\balance
\section{Conclusion}
This paper studied site-specific and frequency-dependent propagation and MIMO performance in the upper mid-band (FR3) using Sionna RT simulations in an indoor laboratory and an outdoor urban scenario. Across representative frequencies spanning FR1, FR3, and FR2, FR3 exhibits an intermediate propagation regime: compared with FR1 it yields more compact delay/angle dispersion in outdoor environments, while compared with FR2, it preserves stronger residual multipath contributions and avoids overly LoS-dominated conditions, which is beneficial for spatial multiplexing. Using channel rank, condition number, and spectral efficiency, we showed that FR3 provides competitive (and often superior) MIMO performance, with clear site dependence driven by environment geometry. Finally, XL-MIMO results indicate that FR3 benefits most under physically meaningful scaling, especially fixed-aperture designs where higher antenna density is realized without the severe propagation limitations of mmWave. These findings support FR3 as a promising band for practical deployments in future 6G systems.

\section*{Acknowledgement}
This work is partly supported by the MultiX and iSEE-6G project under the European Union’s Horizon Europe research and innovation programme (Grant No. 101192521 and 101139291). The work of Zhuangzhuang Cui is supported by the Research Foundation – Flanders (FWO), Senior Postdoctoral Fellowship under Grant No. 12AFN26N.

\balance
\bibliographystyle{IEEEtran}
\bibliography{Reference}

\end{document}